\documentclass[fleqn,usenatbib]{mnras}

\usepackage{newtxtext,newtxmath}

\usepackage[T1]{fontenc}

\DeclareRobustCommand{\VAN}[3]{#2}
\let\VANthebibliography\thebibliography
\def\thebibliography{\DeclareRobustCommand{\VAN}[3]{##3}\VANthebibliography}

\usepackage{graphicx}	
\usepackage{amsmath}	

\usepackage{xcolor}
\usepackage{soul}

\title[]{Characterizing the nucleus of comet 162P/Siding Spring using ground-based photometry}

\author[A. Donaldson et al.]{
A. Donaldson,$^{1}$\thanks{E-mail: a.donaldson@ed.ac.uk}
R. Kokotanekova,$^{1,2,3}$
A. Ro\.zek,$^{1}$
C. Snodgrass,$^{1}$
D. Gardener,$^{1}$
S. F. Green$^{4}$
\newauthor N. Masoumzadeh,$^{5\dagger}$  
and J. Robinson$^{1}$
\\
$^{1}$Institute for Astronomy, University of Edinburgh, Edinburgh EH9 3HJ, UK\\
$^{2}$Bulgarian Academy of Sciences, Institute of Astronomy and National Astronomical Observatory, Sofia 1784, Bulgaria\\
$^{3}$European Southern Observatory, Karl-Schwarzschild-Straße 2, 85748 Garching bei München, Germany \\
$^{4}$School of Physical Sciences, The Open University, Milton Keynes, MK7 6AA, UK\\
$^{5}$Max-Planck-Institut für Sonnensystemforschung, Justus-von-Liebig-Weg 3, 37077 Göttingen, Germany \thanks{now at Deutsches Zentrum für Luft- und Raumfahrt, Institut für Physik der Atmosphäre, Oberpfaffenhofen, Germany} \\
}

\date{Accepted XXX. Received YYY; in original form ZZZ}

\pubyear{2022}

\begin{document}
\label{firstpage}
\pagerange{\pageref{firstpage}--\pageref{lastpage}}
\maketitle

\begin{abstract}

Comet 162P/Siding Spring is a large Jupiter-family comet with extensive archival lightcurve data. We report new $r$-band nucleus lightcurves for this comet, acquired in 2018, 2021 and 2022. 
With the addition of these lightcurves, the phase angles at which the nucleus has been observed range from $0.39^\circ$ to $16.33^\circ$. We absolutely-calibrate the comet lightcurves to $r$-band Pan-STARRS 1 magnitudes, and use these lightcurves to create a convex shape model of the nucleus by convex lightcurve inversion. 
The best-fitting shape model for 162P has axis ratios $a/b=1.56$ and $b/c=2.33$, sidereal period $P=32.864\pm0.001$ h, and a rotation pole oriented towards ecliptic longitude $\lambda_E = 118^{\circ}\pm26^{\circ}$ and latitude $\beta_E = -50^{\circ}\pm21^{\circ}$. We constrain the possible nucleus elongation to lie within $1.4<a/b<2.0$ and discuss tentative evidence that 162P may have a bilobed structure. Using the shape model to correct the lightcurves for rotational effects, we derive a linear phase function with slope $\beta = 0.051\pm0.002$ mag deg$^{-1}$ and intercept $H_r(1,1,0) = 13.86\pm0.02$ for 162P. We find no evidence that the nucleus exhibited an opposition surge at phase angles down to 0.39\textdegree.
 The challenges associated with modelling the shapes of comet nuclei from lightcurves are highlighted, and we comment on the extent to which we anticipate that LSST will alleviate these challenges in the coming decade.

\end{abstract}




\section{Introduction}
\label{intro}

Photometric observations of comets can provide detailed insights into the nature and evolution of these small icy bodies, which are believed to have remained relatively unaltered since their formation 4.6 Gyr ago. Comets with short orbital periods ($<20$ yr) are particularly informative targets, as their properties can be monitored across multiple orbits. Jupiter-family comets (JFCs) are a dynamical subclass of short period comets with perihelia in the inner Solar System and whose orbits are dominated by Jupiter. The central nuclei of these comets are often obscured for observers on the ground by a coma of gas and dust, making it challenging to characterise their physical and surface properties remotely. To date, six short period comets have been imaged by spacecraft: comet 1P/Halley, and five JFCs \citep{snodgrass22missions}. These missions have provided the vast majority of the information we have on the shapes and surface features of comets thus far, demonstrating that their nuclei exhibit a variety of shapes and surface morphologies - from the highly elongated, {\em bilobed} structure of 103P/Hartley 2 \citep{thomas2013shape} to the rounded, asymmetrical appearance of 81P/Wild 2 \citep{brownlee2004surface}.

Of these six comets imaged in situ, four are in highly elongated or bilobed configurations. Halley-type comet 8P/Tuttle made a close passage to Earth in 2008, enabling radar imaging which indicated a bilobate nucleus \citep{harmon2010radar}. With this addition, seven comet nuclei have reliably constrained shapes, and five of these are bilobed - we count 1P/Halley as a bilobed comet, although we note that from the spacecraft images \citep{keller1986first} its shape can only be definitively described as elongated. This sample is too small to provide any statistically significant information on the shapes of the comet population; however, the high fraction observed to be bilobed may indicate that comets have physical characteristics or undergo formation and/or evolution processes that result preferentially in these shapes. Possibilities for the formation of bilobate nuclei have been explored at various dynamical stages: for example, slow growth by hierarchical agglomeration in the primordial disk \citep{davidsson2016}; re-accretion following shape-changing collisions \citep{2017jutzi, 2018schwartz, bagatin2020gravitational}; slow collisions between bodies of similar size \citep{2015jutzi}; and sublimation-driven disruption in the Centaur region \citep{safrit2021subtorques}. Bilobed shapes are also found in other small body populations. The New Horizons target 486958 Arrokoth (2014 MU$_{69}$), a $\sim$30 km body in the Cold Classical population of trans-neptunian objects, appears to be comprised of two distinct, flattened lobes \citep{stern2019arrokoth}. In the near-Earth asteroid (NEA) populations, the contact binary fraction is estimated to be between 15 and 30 per cent \citep{virkkibinary}.

While our current understanding of the shapes of comet nuclei is dominated by information returned from in situ missions, it is also possible to place constraints on shape information using ground-based observations. Radar facilities can be used when comets make close approaches to Earth, as was the case for comet 8P. When it is possible, delay-Doppler imaging is capable of providing nucleus size, rotation and shape information \citep{harmon2004comets}. In extremely rare circumstances, a comet can pass in front of a star in such a way that its nucleus properties can be constrained by stellar occultation methods (discussed in e.g. \citeauthor{fernandez1999inner} \citeyear{fernandez1999inner}). This method relies on observers in multiple geographic locations and precise ephemerides for advance planning. However, due to the unpredictable nature of cometary orbits as they are affected by activity-driven torques, this technique has not yet successfully provided shape information for any comet nucleus.

The most readily available means of acquiring shape information is from rotational lightcurves. The shape of the lightcurve is dependent primarily on the changing projected cross-section of the irregularly-shaped nucleus as it rotates. {\em Convex lightcurve inversion} (CLI) is a technique that can be used to extract shape information from lightcurves \citep{kaasa1, kaasa2}, provided they cover a wide range of observing geometries. The CLI procedure generates a convex hull representing the object's global shape, `inverting' the lightcurves to create a shape that is capable of reproducing the lightcurve at each input observing geometry and rotational phase. CLI is regularly employed to produce convex shape models of asteroids from their photometric disk-integrated lightcurves. The shape models for more than 3000 asteroids are stored in the Database of Asteroid Models from Inversion Techniques (DAMIT\footnote{https://astro.troja.mff.cuni.cz/projects/DAMIT}; \citealt{durechdamit}), of which a large number have been produced using CLI.

In contrast, only one comet nucleus has been modelled by CLI to date\footnote{Two unusual objects which are related to JFCs have also had their shapes modelled by lightcurve inversion: (3552) Don Quixote and 323P/SOHO}. This was the target of ESA's Rosetta mission, comet 67P/Churyumov-Gerasimenko (hereafter 67P), prior to the rendezvous phase of the spacecraft \citep{lowry2012nucleus}. Using nucleus lightcurves collected around the 2006 aphelion passage (at heliocentric distances r$_h>$3 AU) and nucleus images from HST at r$_h\sim2.5$ AU, the authors produced a convex model. The CLI model was later refined with the addition of lightcurves from the Rosetta approach \citep{mottola2014rotation}, which yielded a shape with properties similar to the original - consistent elongation and the presence of large planar surfaces. Both models produced pole orientation consistent with those found later by the spacecraft \citep{preusker2015shape}, and the lightcurve data was used to detect a period change for 67P between the two perihelion passages. A direct comparison of the convex and Rosetta shape models can be found in \cite{snodgrass22missions}. By definition, CLI cannot reproduce large-scale concave features. The substantial flat regions on the shape model of 67P masked what Rosetta observed to be a slim neck connecting two lobes of a rounded, bilobed shape. Large flat facets in convex models are strong indicators of large scale surface concavities \citep{devogele2015}. Physical considerations regarding the criteria for the stability of any shape model may also be used to infer a bilobed shape, given that it is unlikely that single-lobed objects with extremely elongated axis ratios (a/b $>$ 2.3) would have formed naturally \citep{mcneill2018, jeans1919problems}, though 67P has proven that bilobed shapes can be compact and rounded rather than elongated.

In this work, we apply CLI to comet 162P/Siding Spring (hereafter 162P). The most recent physical properties measured for this comet are outlined in Table \ref{tab:162p}. 162P is a particularly large JFC, with effective radius $R=7.03^{+0.47}_{-0.48}$ km \citep{fernandez2013thermal}. Estimates of its geometric albedo ($p_R$) suggest it has one of the darkest surfaces of all studied comets. The comet has a relatively long synodic rotation period ($P_s$) around $32.88$ h, and has presented no evidence for significant period changes between orbits  to date \citep{2018kokotanekova}. Initially discovered as asteroid 2004 TU12, it was recharacterised as a comet after it was observed to display intermittent activity when closest to the Sun \citep{campins2006nuclear}. These infrequent displays of weak activity may suggest that the comet is approaching dormancy, as defined by \cite{kresakdormant} and \cite{hartmanndormant}. The lack of a detectable coma over most of the comet's orbit has allowed for good lightcurve coverage of the nucleus at a range of viewing geometries, making 162P an appealing choice for this study. We obtained new lightcurves of the nucleus of 162P between 2018-2022, presented in Section \ref{odr} along with a description of the absolute photometric calibration process. The application of convex lightcurve inversion to our entire set of nucleus lightcurves is discussed in Section \ref{methods}, including all results for 162P. The physical properties of the nucleus that we can derive from the shape model and a more general outlook on the possibilities of convex inversion for the lightcurves of comet nuclei are outlined in Section \ref{disc}.

\begin{table}
 \centering
 \caption{Summary of the physical properties of comet 162P. \newline
 $^*$The value given for the period is the best-fit value based on Monte Carlo trials combining datasets at different observing geometries. The range of possible values for the synodic rotation period is $32.812 - 32.903$ h.} 
 \label{tab:162p}
 
 \begin{tabular}{|l|c|c|}
  \hline
   Property (symbol) & Value & Reference \\
  \hline
    Orbital period & 5.46 y & \cite{horizons} \\
    Inclination ($i$) & $27.5^\circ$  & \cite{horizons} \\
    Rotation period ($P_s$) & 32.877$^{+0.026}_{-0.065}$ h$^*$ & \cite{2018kokotanekova} \\
    Geometric albedo (p$_R$) & $0.022\pm0.003$ & \cite{2017kokotanekova} \\
    Effective radius ($R$) & 7.03$^{+0.47}_{-0.48} $ km & \cite{fernandez2013thermal} \\

  \hline
 \end{tabular}
\end{table}


\section{New observations and data reduction}
\label{odr}

\subsection{Observations}
\label{obs}

In this work we have combined all published optical lightcurves of the nucleus of 162P/Siding Spring collected between 2007-2017 with new lightcurves obtained in 2018, 2021 and 2022. A complete summary of all the lightcurves is given in Table \ref{tab:obs}. Lightcurves collected prior to 2018 (ID 1-12 in Table \ref{tab:obs}) have been reported previously in \citet{2017kokotanekova} and \citet{2018kokotanekova}. The new lightcurves were obtained in April 2018 (ID 13-15), July 2018 (ID 16-18), December 2021 (19-21), January 2022 (ID 22-28) and March 2022 (ID 29-33), and were collected and processed in this work as described below.

\begin{table*}

 \caption{Summary of complete set of 162P observations analysed in this work. The lightcurves from 2007-2017 (IDs 1-12) have been published previously in \citet{2017kokotanekova} and \citet{2018kokotanekova}. The observations and calibration procedure described in this work apply to the 2018-2022 lightcurves, IDs 13-33.}
 \begin{tabular*}{0.95\textwidth}{@{}l@{\hspace*{15pt}}l@{\hspace*{15pt}}l@{\hspace*{15pt}}l@{\hspace*{15pt}}l@{\hspace*{15pt}}l@{\hspace*{15pt}}l@{\hspace*{15pt}}l@{\hspace*{15pt}}l@{\hspace*{15pt}}l@{}}
  \hline
  ID & UT Start date & Instrument & Filters & Exposure time [s] & $r_h$ [au] & $\Delta$ [au] & $\alpha$ [deg.] & $\lambda_E$ [deg.] & $\beta_E$ [deg.] \\
  \hline
  1 & 2007-05-17  & WHT/PFIP & R & 13$\times$90 & 4.86 & 4.03 & 7.51 & 205.3 & 3.1 \\
  2 & 2007-05-18  & WHT/PFIP & R & 3$\times$90, 10$\times$110 & 4.86 & 4.04 & 7.69 & 205.4 & 3.1 \\
  3 & 2007-05-19  & WHT/PFIP & R & 12$\times$90 & 4.86 & 4.05 & 7.86 & 205.4 & 3.1 \\ 
  4 & 2012-04-23  & VLT/FORS2 & R & 30$\times$60 & 4.73 & 3.79 & 4.68 & 197.5 & 7.1 \\ 
  5 & 2012-05-24  & VLT/FORS2 & R & 5$\times$60 & 4.77 & 4.12 & 10.02 & 199.2 & 6.3 \\ 
  6 & 2012-06-14  & NTT/EFOSC2 & R & 18$\times$180 & 4.80 & 4.44 & 11.84 & 200.4 & 5.7 \\ 
  7 & 2012-06-17  & NTT/EFOSC2 & R & 13$\times$300 & 4.80 & 4.49 & 11.97 & 200.6 & 5.6 \\
  8 & 2012-06-23  & VLT/FORS2 & R & 29$\times$60 & 4.81 & 4.59 & 12.14 & 200.8 & 5.4 \\
  9 & 2017-02-17  & INT/WFC & $r'$ & 93$\times$120 & 4.30 & 3.58 & 9.88 & 185.8 & 12.7 \\
  10 & 2017-02-18  & INT/WFC & $r'$ & 52$\times$120 & 4.31 & 3.57 & 9.71 & 185.9 & 12.7 \\
  11 & 2017-02-21  & INT/WFC & $r'$ & 43$\times$120, 36$\times$150 & 4.31 & 3.55 & 9.18 & 186.1 & 12.6 \\
  12 & 2017-02-26  & Rozhen 2m/FoReRo & R & 21$\times$300 & 4.33 & 3.51 & 8.24 & 186.5 & 12.4 \\
  13 & 2018-04-10  & INT/WFC & $r'$ & 80$\times$90 & 4.89 & 3.90 & 2.26 & 209.6 & 0.8 \\
  14 & 2018-04-11 & INT/WFC & $r'$ & 71$\times$90 & 4.89 & 3.90 & 2.02 & 209.6 & 0.8 \\
  15 & 2018-04-18 & INT/WFC & $r'$ & 6$\times$120, 41$\times$200 & 4.89 & 3.89 & 0.39 & 210.0 & 0.6  \\
  16 & 2018-07-09 & NTT/EFOSC2 & $r$ & 9$\times$240, 4$\times$200, 3$\times$180 & 4.85 & 4.66 & 12.06 & 214.2 & -1.7 \\
  17 & 2018-07-10 & NTT/EFOSC2 & $r$ & 2$\times$180, 6$\times$300, 6$\times$360 & 4.85 & 4.67 & 12.08 & 214.3 & -1.7 \\
  18 & 2018-07-11 & NTT/EFOSC2 & $r$ & 2$\times$200, 5$\times$300& 4.85 & 4.69 & 12.09 & 214.4 & -1.7 \\
  19 & 2021-12-29 & LT/IO:O & $r'$ & 15$\times$150 & 3.49 & 3.29 & 16.33 & 166.9 & 19.9 \\
  20 & 2021-12-30 & LT/IO:O & $r'$ & 6$\times$150 & 3.50 & 3.28 & 16.29 & 167.1 & 19.9 \\
  21 & 2021-12-31 & LT/IO:O & $r'$ & 3$\times$150 & 3.50 & 3.27 & 16.24 & 167.2 & 19.8 \\
  22 & 2022-01-01 & LT/IO:O & $r'$ & 9$\times$150 & 3.51 & 3.26 & 16.19 & 167.3 & 19.8 \\
  23 & 2022-01-03 & LT/IO:O & $r'$ & 8$\times$180 & 3.52 & 3.24 & 16.07 & 167.5 & 19.7 \\
  24 & 2022-01-04 & LT/IO:O & $r'$ & 7$\times$180, 7$\times$175 & 3.52 & 3.23 & 16.01 & 167.6 & 19.7 \\
  25 & 2022-01-05 & LT/IO:O & $r'$ & 9$\times$180 & 3.52 & 3.22 & 15.95 &167.7 & 19.6 \\
  26 & 2022-01-06 & LT/IO:O & $r'$ & 8$\times$180, 7$\times$175 & 3.53 & 3.21 & 15.88 & 167.9 & 19.6 \\
  27 & 2022-01-07 & LT/IO:O & $r'$ & 5$\times$180, 4$\times$175 & 3.53 & 3.20 & 15.80 & 168.0 & 19.6 \\
  28 & 2022-01-11 & LT/IO:O & $r'$ & 10$\times$180 & 3.55 & 3.16 & 15.46 & 168.4 & 19.4 \\
  29 & 2022-03-03 & INT/WFC & $r'$ & 20$\times$60 & 3.79 & 2.88 & 6.86 & 173.7 & 17.5 \\
  30 & 2022-03-04 & INT/WFC & $r'$ & 45$\times$60 & 3.79 & 2.88 & 6.72 & 173.8 & 17.5 \\
  31 & 2022-03-05 & INT/WFC & $r'$ & 52$\times$60 & 3.79 & 2.88 & 6.58 & 173.9 & 17.4 \\
  32 & 2022-03-06 & INT/WFC & $r'$ & 65$\times$60 & 3.80 & 2.88 & 6.44 & 174.0 & 17.4 \\
  33 & 2022-03-08 & INT/WFC & $g'$, $r'$ & 20$\times$60 $(r')$, 18$\times$60 $(g')$ & 3.80 & 2.88 & 6.21 & 174.1 & 17.3 \\
  
  \hline
  \label{tab:obs}
 \end{tabular*}
\end{table*}


For all observations, the field containing the comet was tracked at a sidereal rate, to enable the background stars and comet to be treated as point sources in the aperture photometry procedure described in Section \ref{abscal}. The duration of each exposure was chosen such that the comet did not trail by more than the radius of the seeing disc. The April 2018 and March 2022 frames were obtained using the wide-field camera (WFC) on the 2.5-m Isaac Newton Telescope (INT) situated at Roque de los Muchachos Observatory (ORM), La Palma, Spain. The WFC is made up of four $2048\times 4100$ pixel CCDs, with a combined field of view (FoV) $34\times 34$ arcmin. The pixel scale of each CCD is 0.33 arcsec/pixel in $1\times 1$ binning mode. The target was kept in CCD4 throughout each epoch, and was imaged in the WFCSloanR filter in 2018, and in both WFCSloanR $(r')$ and WFCSloanG $(g')$ in 2022. The mean seeing varied nightly between $1.2-2$" for the 2018 observations, and $1.3-1.5$" in 2022. Bias frames were obtained nightly and used to subtract the bias level from all frames. Twilight flats in both filters were obtained when possible at the beginning of each night. These were used to flat-field correct frames taken on the same night (on nights when no twilight flats were obtained, frames were corrected using flats from the previous or next night).

The July 2018 frames were collected with the European Southern Observatory (ESO) New Technology Telescope (NTT) at La Silla Observatory, Chile, using the ESO Faint Object Spectrograph and Camera (EFOSC). This is a multi-mode instrument with a $4.1 \times 4.1$ arcmin FoV. The 162P frames were obtained using the Gunn-r ($r$) filter with the detector in $2\times 2$ binning mode, providing an effective pixel scale of 0.24 arcsec/pixel. The nightly seeing was around 1.4". The bias subtraction and flat-field correction were applied to the frames by the same method as for the INT frames above.

The December 2021/January 2022 photometry was acquired with the optical imaging component of the Infrared-Optical (IO:O) instrument on the fully robotic 2.0-m Liverpool Telescope (LT) at ORM. The IO:O is comprised of a $4096\times4112$ pixel CCD, with an unbinned pixel scale of approximately 0.15 arcsec/pixel and effective FoV $10\times10$ arcmin. The data were obtained in $2\times2$ binning mode, using the SDSS-$r'$ filter. Photometry was collected in short nightly observing blocks (OB) over a ~2-week period ($\sim$5-25 exposures per OB) to probe as much of the object's long rotation period as possible. The average nightly FWHM of the seeing disc was $\sim$2". The frames were bias subtracted and flat-field corrected automatically in the IO:O data reduction pipeline.

\begin{figure}
    \centering
    \includegraphics[width=0.98\linewidth]{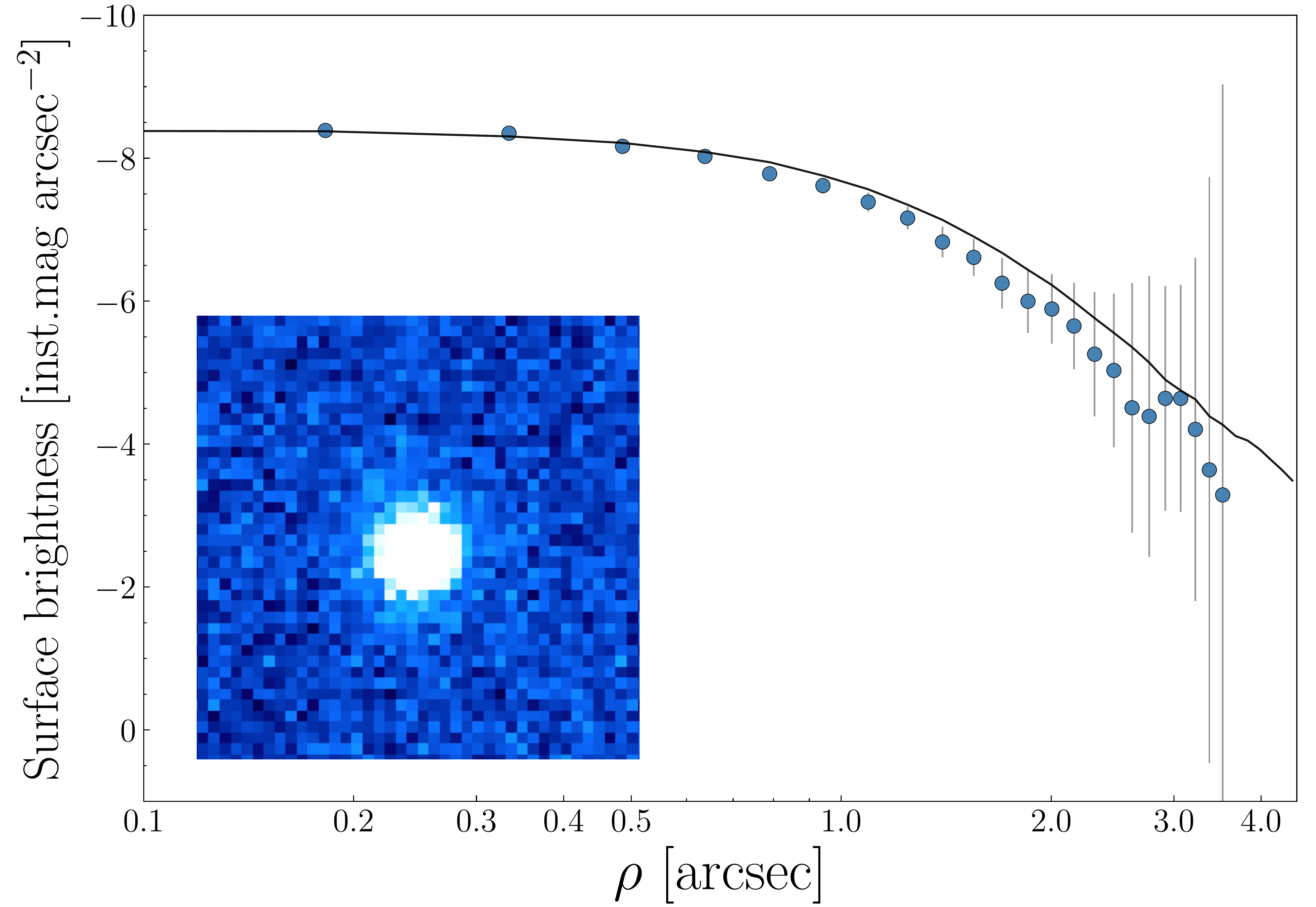}
    \caption{Surface brightness profile of 162P in frames obtained on 01-01-2022. Black line shows radial profile of a field star, scaled to match the surface brightness of the comet within its smallest aperture. The comet profile and stellar profiles are indistinguishable within the uncertainties in the comet's surface brightness. Inset shows the composite background-subtracted comet.}
    \label{fig:psf}
\end{figure}

To minimize the possibility of the nucleus signal being contaminated by coma, the comet was observed at heliocentric distances $r_h>3$.5 au. We confirmed that 162P was inactive by aligning and stacking background-subtracted comet frames, and comparing the radial profile of the comet stack to the PSF of a scaled stacked comparison star for each epoch of observations. An example is shown for the LT observations in Fig. \ref{fig:psf}. The comet's radial profile is point source-like with no evidence of extended brightness at large distances from the centre, indicating that no detectable activity was present at the time of the observations.

\subsection{Absolute calibration procedure}
\label{abscal}

To combine the multi-epoch photometry obtained with multiple instruments, we followed the procedure outlined in Section 3.4 of \citet{2017kokotanekova} with some alterations to the software used. This allowed us to transform differential photometry of the comet nucleus over the course of a night into absolutely-calibrated magnitudes using the $r$-band magnitudes of background stars from the Pan-STARRS 1 (PS1) catalogue \citep{chambersps1}. We used \textsc{aperture\textunderscore photometry} and related routines from the Astropy \textsc{Photutils} package \citep{photutils} to extract instrumental magnitudes, and \textsc{calviacat} \citep{calviacat} to calibrate the photometry of the comet using comparison stars on the frames.

For each night of observations, we identified background sources that appear on all frames. We then cross-checked each source with the PS1 catalogue using \textsc{calviacat}, and used only those with catalogue entries. We further filtered these sources to remove potential galaxies, sources that are too close together for reliable background subtraction, and sources with extreme colour indices compared to solar i.e $(g-r) < 0$ and $(g-r)>1.5$. For the NTT frames we found a large number of background stars with $(g-r)> \sim$1.1, and so further limited the colour indices of the comparison stars to within $0<(g-r)<1.1$ to prevent biasing the colour towards more extreme values.
    
  We performed aperture photometry of the final background stars and the comet, using the suite of tools included in \textsc{Photutils}. For the photometry, we used a circular aperture with radius equal to the median FWHM of the background stars on each frame. This choice of aperture size was found to maximise the signal-to-noise ratio (SNR) of the comet. The number of comparison stars used for photometry was dependent on the field each night, and varied between $\sim$10-100 stars per night. For the INT frames, we corrected the background-subtracted aperture fluxes for the effects of distortion caused by distance from the optical axis of the wide-field instrument, as described in \cite{gonzalez2008initial}, before converting them to magnitudes.
    
 To absolutely calibrate the lightcurve points, it is necessary to account for the different colour responses of each instrument. We determined the colour term ($CT$) for each instrument following the iterative-gradient method detailed in \citet{2017kokotanekova}. The values obtained for the CTs are given in Table \ref{tab:CTs}. For the LT $r'$ frames and INT $g'$ frames, we used the final gradient values as the CTs by which to correct our photometry. The CTs for the INT/WFC $r'$ and NTT/EFOSC $r$ filters were previously derived using a much larger sample of comparison stars by the works stated in Table \ref{tab:CTs}. We therefore ensured that their CTs provided suitable linear fits to our background stars, and used these values to correct for the $r$-band colour response of the INT/WFC and NTT/EFOSC. 

\begin{table}
 \centering
 \caption{The colour terms (CT) used in the absolute-calibration process to account for the varying colour response of each instrument/filter.}
 \label{tab:CTs}
 \begin{tabular*}{\linewidth}{llll}
  \hline
  Instrument & Filter & CT & Reference\\
  \hline
  INT/WFC & $g'$ & $-0.044\pm0.015$ & This work \\
  INT/WFC & $r'$ & $0.008\pm0.004$ & \cite{2018rkphd} \\
  NTT/EFOSC & $r$ & $-0.194\pm0.005$ & \cite{2017kokotanekova} \\
  LT/IO:O & $r'$ & $0.000\pm0.005$ & This work \\

  \hline
 \end{tabular*}
\end{table}

 In order to determine the final $r_{PS1}$ magnitude for the nucleus, we required an estimate of its $(g-r)_{PS1}$ colour. We obtained this using the $g$ and $r$ frames from lightcurve ID 33. The observations on this night were taken in single-filter blocks of ten frames as $rgrg$, allowing us to infer the average $r$ magnitude at the time of the $g$ observations and vice versa. This gave an average $(g-r)_{PS1}$ nucleus colour of $0.48\pm0.04$. This is consistent with value for $(B-V)$ of $0.76\pm0.01$ determined by \cite{lamy2009colors} for 162P: converting this to $(g-r)_{PS1}$ using the expressions and coefficients given in \cite{tonry2012pan} yields $0.51\pm0.03$.

The lightcurve IDs 1-12 reported in \citet{2018kokotanekova} were calibrated using an average JFC surface colour index $(g-r)_{PS1} = 0.58\pm 0.06$. These lightcurves were reprocessed following the procedures outlined in their work, but using the updated $(g-r)_{PS1}$ nucleus colour index found in this work to ensure consistent calibration.
 
We corrected the calibrated lightcurves to account for the changing heliocentric ($r_h$) and geocentric ($\Delta$) distances and solar phase angle ($\alpha$) of the comet, to combine all lightcurve points to a consistent orbital configuration. The absolute magnitude H$_{r,PS1}(1,1,\alpha=0)$ is given by Equation \ref{eq:reduced} and represents the magnitude of the comet at a theoretical position at distance 1 au from both the Sun and Earth and solar phase angle $\alpha=0$\textdegree, assuming a linear phase function $\beta$. 
   
   \begin{equation}
       \label{eq:reduced}
       H_{r,PS1}(1,1,\alpha=0) = m_r - 5\log_{10}(R_h \Delta) - \beta \alpha
   \end{equation}

\begin{figure}
    \centering
    \includegraphics{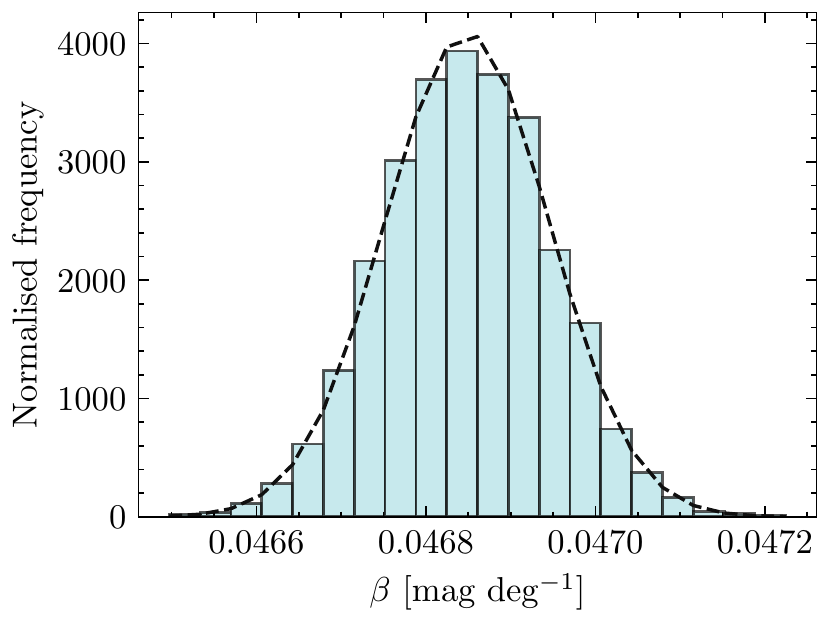}
    \caption{Distribution of phase function ($\beta$) values resulting from 5000 randomised lightcurve trials. The distribution has mean $\beta=0.0468\pm0.0001$ mag deg$^{-1}$.}
    \label{fig:betamc}
\end{figure}

\begin{figure}
\centering
    \includegraphics[trim=0cm 0.0cm 0cm 0cm, width=\linewidth]{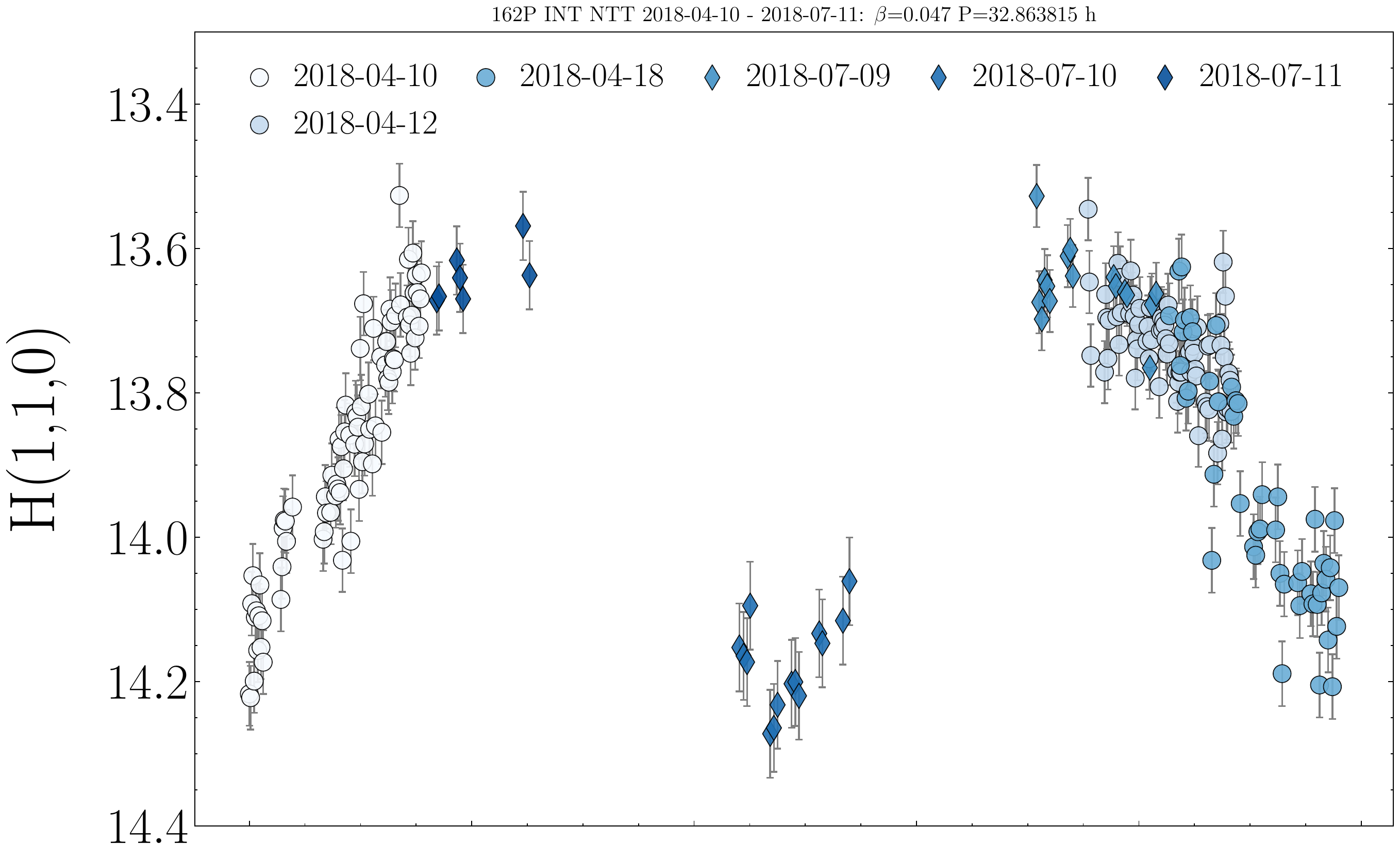}
    \includegraphics[width=\linewidth]{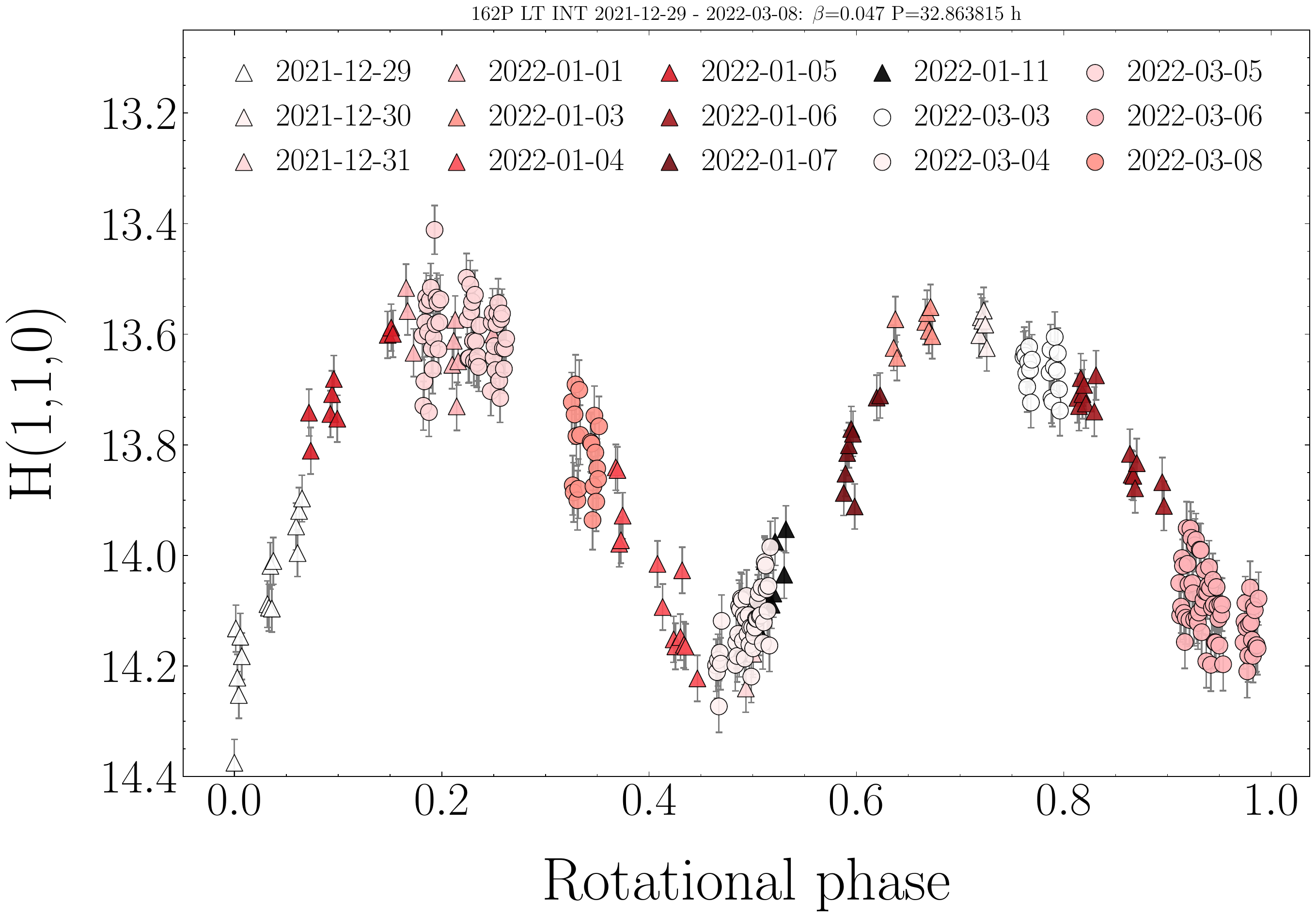}

\caption{Lightcurves for 162P using photometry obtained in 2018 (upper) and 2021-2022 (lower). The comet magnitude at each point has been absolutely-calibrated to the Pan-STARRS 1 $r$ filter. The diamond-shaped lightcurve points were observed at the NTT, circular points with the INT, and triangular points with the LT. All points have been corrected for phase function using the best fit value for $\beta=0.0468$ mag deg$^{-1}$ resulting from the Monte Carlo method described in the text. The lightcurves have been phased to a period of 32.8638 h, the best-fit sidereal period for the comet according to the convex inversion procedure described in Section \ref{methods}.}

\label{fig:lcs}
\end{figure}

Following the method of \cite{2017kokotanekova}, we fit the phase function $\beta$ by a Monte Carlo (MC) method which accounts for the photometric uncertainty of each lightcurve point. We replaced each lightcurve point with a value drawn randomly from a normal distribution with mean equal to the magnitude value and standard deviation equal to the uncertainty on that lightcurve point. We repeated this to produce 5000 randomised lightcurves, and used linear regression to determine the best fitting linear phase function for each random lightcurve. We plotted the resulting $\beta$ distribution and fit a Gaussian function to it, as shown in Fig. \ref{fig:betamc}. We select the best fit value of $\beta$ and its corresponding uncertainty as the mean and standard deviation of this function, $\beta=0.0468\pm0.0001$ mag deg$^{-1}$. Fig. \ref{fig:lcs} shows the final absolute lightcurves from 2018 and 2021-2022, both phased to a period of 32.8638 h (the best fitting rotation period derived in this work, see Section \ref{methods}). We measure a lightcurve amplitude $\Delta m$ for the 2021-2022 points of approximately $0.75$ (not including the single outlying points at 2021-12-29 and 2022-03-05), though this is likely an overestimate due to the large amount of scatter in the lightcurves around the peak at rotational phase $\sim$0.2.

\section{Modelling the nucleus shape}
\label{methods}

\subsection{Convex lightcurve inversion}
With the absolutely-calibrated time series photometry at hand, we constructed the shape model of 162P by convex lightcurve inversion (CLI). The inversion procedure used in this work is the publicly available \textsc{convexinv} package by \cite{durechdamit}, adapting the algorithms described by \cite{kaasa1} and \cite{kaasa2}. The CLI software models the shape, sidereal period and rotation pole orientation that best fit the input lightcurves. The model is produced using a Levenberg-Marquardt algorithm \citep{press1992numerical} to optimise the area and orientation each facet of a convex polyhedron by fitting a number of shape-related parameters. For a given combination of fitting parameters, the software computes a relative value of $\chi^2$ that quantifies how well the shape matches the observed lightcurve points \cite[as defined in][]{kaasa1}. The combination of parameters corresponding to the smallest value of $\chi^2$ are referred to as the `best-fit' parameters. 

CLI is most effective when the object of interest is observed at a large number of viewing angles or {\em observing geometries}. This term refers to the object's position and orientation relative to both the observer on Earth and to the Sun. Changing the observing geometry changes the fraction of the object's surface that is illuminated at a given time, which in turn affects the reflected flux measured by the observer. This effect is compounded by the shape and rotation state of the object. If the aspect angle (the angle subtending the rotation pole and the observer's line of sight) changes significantly between observations, then the projected cross-section and brightness of the object will change, assuming it is non-spherical. For example, a tri-axial ellipsoid rotating in a stable configuration about its shortest axis will have a maximum projected cross-sectional area when its aspect angle is equal to 0\textdegree\space i.e. when the line of sight of the observer is perpendicular to the rotation axis. When lightcurves sample a wide range of different observing geometries, the projected cross-section of the object is viewed from many different orientations as the relative position of the object, observer, and Sun change. This allows for a well-constrained shape model. The solar phase angle $\alpha$ also plays a role in the quality of the final model: increased shadowing effects at large $\alpha$ can help constrain information about the shape that is not possible at small $\alpha$. For this reason, it is ideal for convex inversion to have access to lightcurves at large solar phase angles $\sim$20\textdegree\space and above \citep{kaasa1}.

The CLI steps are described in detail below. A key aspect of the \textsc{convexinv} software is that input lightcurves can be treated as either {\em relative} or {\em calibrated}. The software defines calibrated lightcurves as those that have been brought to the same magnitude scale at a unit distance of 1 au from the Earth and from the Sun. This corresponds to the form of the lightcurves produced in Section \ref{abscal} {\em without} the phase function correction. Treating the lightcurves as calibrated effectively means that the shape optimisation procedure does not shift lightcurves from different epochs with respect to one another in the fitting process. In this case, to account for the effects of the changing solar phase angle between lightcurves, the software fits a light-scattering function during the shape optimisation procedure that includes a phase function. Rather than assuming an explicit scattering law, \textsc{convexinv} fits an empirical function, described in detail in \cite{kaasa2}. 
The built-in phase function $f(\alpha)$ used by this scattering function to incorporate the effects of solar phase on the observed lightcurve points into the shape model is given by Equation \ref{eq:pf}:

\begin{equation}
\label{eq:pf}
f(\alpha) = a_s \exp(-\frac{\alpha}{d_s})+k_s\alpha+1
\end{equation}

The phase function is parametrized by $a_s, d_s$ and $k_s$, which are fitted in the convex inversion procedure, and $\alpha$ is the solar phase angle of the lightcurve points. It is important to note that $f(\alpha)$ and the linear phase function $\beta$ described in Section \ref{abscal} are distinct. The \textsc{convexinv} software requires that the lightcurve points be heliocentric and geocentric-distance corrected and converted to intensity space. These intensities are not corrected for phase angle: determining the phase function correction is part of the shape optimisation procedure through fitting $a_s,d_s$ and $k_s$. To produce the shape model, we chose to implement and fit a phase function that is linearly dependent on phase angle, (i.e. fit a value solely for $k_s$ to convergence during the shape modelling procedure, fixing $a_s$ and $d_s$ at 0 and 1 respectively). We justify this choice below in Section \ref{testpf}. We stress however that no physical significance is placed on the final fitted values of these parameters, particularly $k_s$: we treat this as purely empirical. Identifying any relationship between the best-fit value for $k_s$, defined in intensity space and $\beta$, defined in magnitude space, is not necessary for this analysis.

\subsection{Period search}
To find the best shape, it is necessary to have a good measurement of the object's sidereal period. We used the \textsc{periodscan} procedure included in the convex inversion software package to find the best-fitting sidereal period for 162P. \textsc{periodscan} trials period values incrementally within a user-defined range. The range is typically based on prior knowledge of the object's rotation period because \textsc{periodscan} is computationally intensive, particularly for short rotation periods. Since the period of 162P was previously estimated to be $\sim$32.9 h, we used trial periods in the range 10-70 h. This ensured that periods around half and twice the literature estimate were also tested.
For each trial period, \textsc{periodscan} starts from six initial pole locations and calls \textsc{convexinv} to simultaneously optimise the shape and scattering function. The relative $\chi^2$ value of the optimised shape at that trial period is determined and output - no other information about the shape or spin is stored at this stage.

\begin{figure}
\centering
    \includegraphics[width=\linewidth]{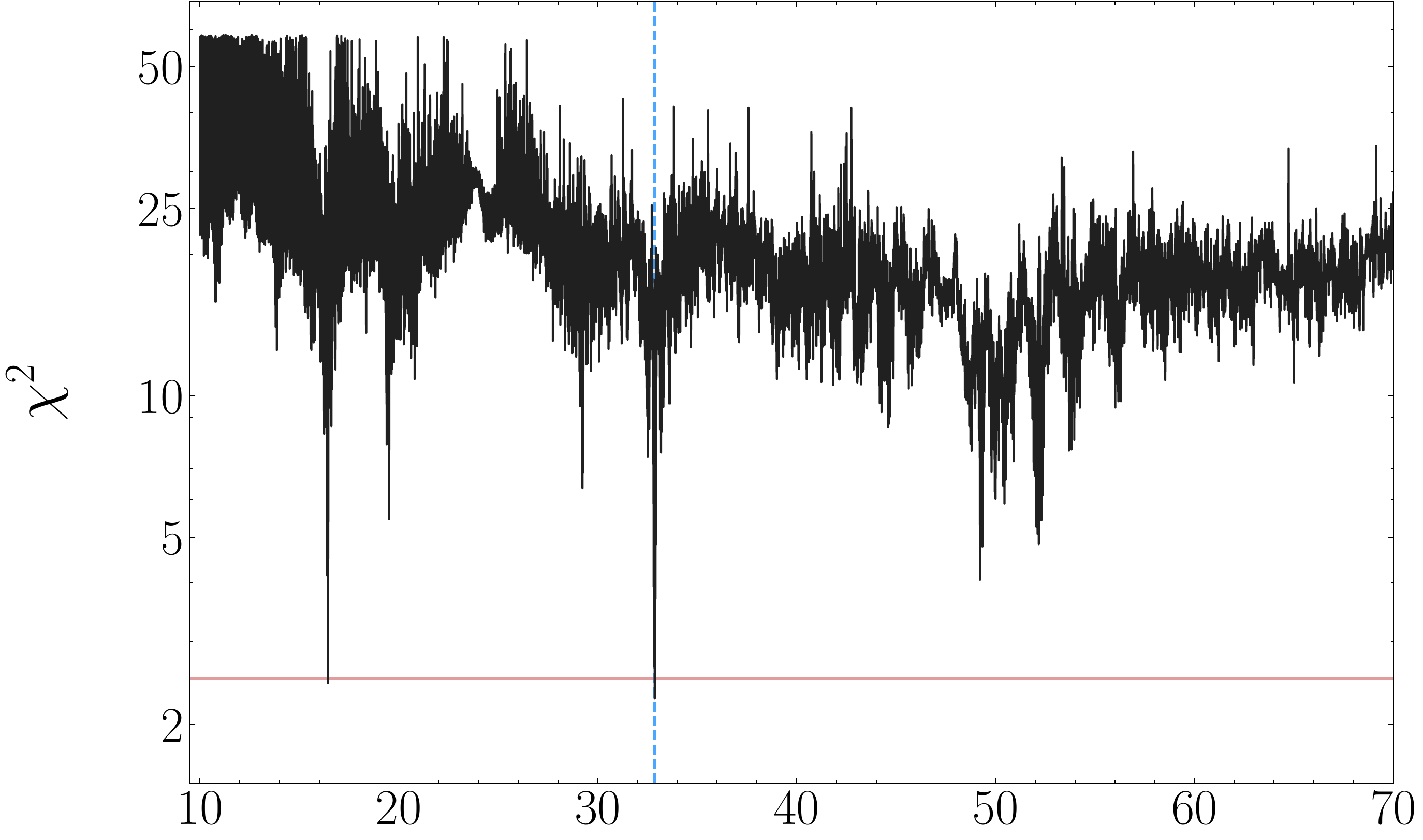}
   
    \includegraphics[width=\linewidth]{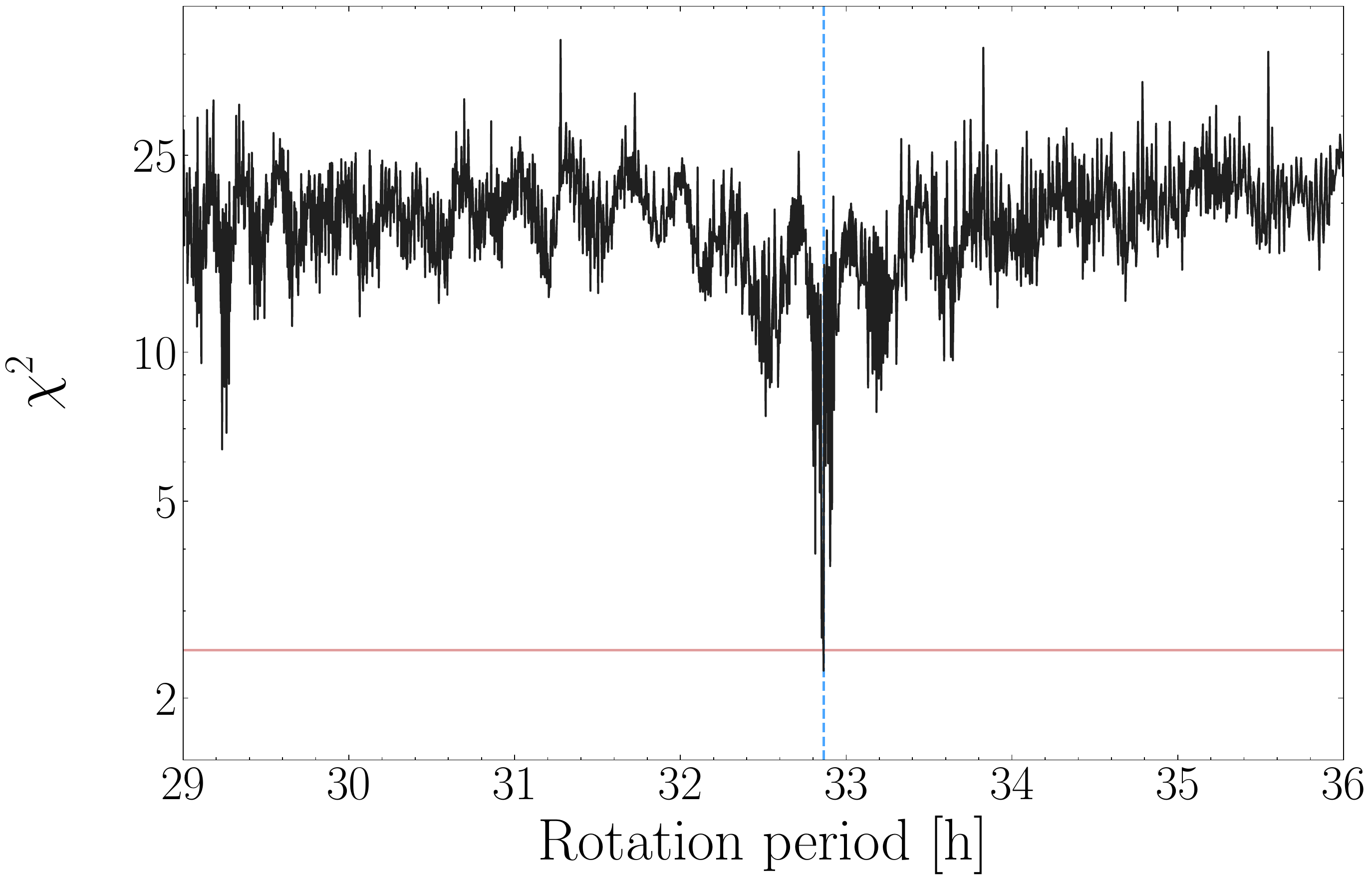}

\caption{Periodograms demonstrating the quality of the sidereal period fit at each trial period using the \textsc{periodscan} routine over an interval of $10-70$ h (upper) and zooming in on interval $29-36$ h (lower). The dashed vertical line marks the period corresponding to the lowest value of $\chi^2$, $P=32.8638$ h, and the solid horizontal line shows the 10 per cent increase above the minimum $\chi^2$. }
\label{fig:period}

\end{figure}

The resulting periodogram is shown in Fig. \ref{fig:period}. The period corresponding to the minimum value of $\chi^2$ is $P=32.8638$ h, which is close to the literature synodic period.
The next lowest $\chi^2$ value corresponds to $P=16.4319$h, half the value of the best fit period. We rule out this solution for the period as it is unlikely that the rotation of an irregularly-shaped object would produce a single-peaked lightcurve. There are several additional global peaks at $P=19.5079$ h,  $P=29.2353$ h, $P=49.2190$ h and $P=52.1698$ h. We ruled these out as possibilities for the sidereal period by plotting the absolutely-calibrated lightcurve points shown in Fig. \ref{fig:lcs} phased to each of these periods in turn, none of which resulted in a coherent double-peaked lightcurve for either of the 2018 or 2021-2022 datasets.

To associate a formal uncertainty with the best fit period value by the method described in \cite{durech2012analysis} leads to a period range defined by an increase of 5 per cent of the minimum $\chi^2$ value (assuming 800 degrees of freedom, from $\sim$900 individual lightcurve points and $\sim$100 model parameters). However, there are only minuscule differences in the models produced by period values within 10 per cent of the minimum $\chi^2$ value \citep{rozek2022}, and as such we use this limit to express the uncertainty in the period. We therefore give the value of the best fit sidereal period as $P=32.8638\pm0.0007$ h. 

For this analysis, we assume that the sidereal period is constant between observing epochs. This assumption is not unrealistic: 162P is a particularly large JFC and displays only very weak levels of outgassing, implying that it would experience negligible changes to its rotation period between orbits \citep{samarasinha2013}. The synodic period of 162P was previously estimated from the lightcurves acquired in 2007, 2012 and 2017 (ID 1-12) by \citet{2017kokotanekova,2018kokotanekova}. The authors found that periods in the range $32.812-32.903$ h provided good fits to the combined lightcurves from all three epochs. By measuring the change in the phase angle bisector for each epoch, we determined synodic periods of 32.8856 h, 32.8794 h and 32.8866 h for the 2007, 2012 and 2017 epochs respectively, assuming a constant sidereal period $P=32.8638$ h. These values are all within the range of common synodic period values identified by \citet{2018kokotanekova}, implying that the sidereal period measured in this work is in agreement with these previously-estimated synodic periods. Additionally, the small deviations between the best individual synodic period fits for each of the 2007, 2012 and 2017 observing epochs in \citet{2017kokotanekova, 2018kokotanekova} can be attributed to the comet's varying observing geometry.

\subsection{Optimising shape and spin state}

\begin{figure}
    \centering
    \includegraphics[trim=4.5cm 7.6cm 4cm 8cm, clip=true, width=\linewidth]{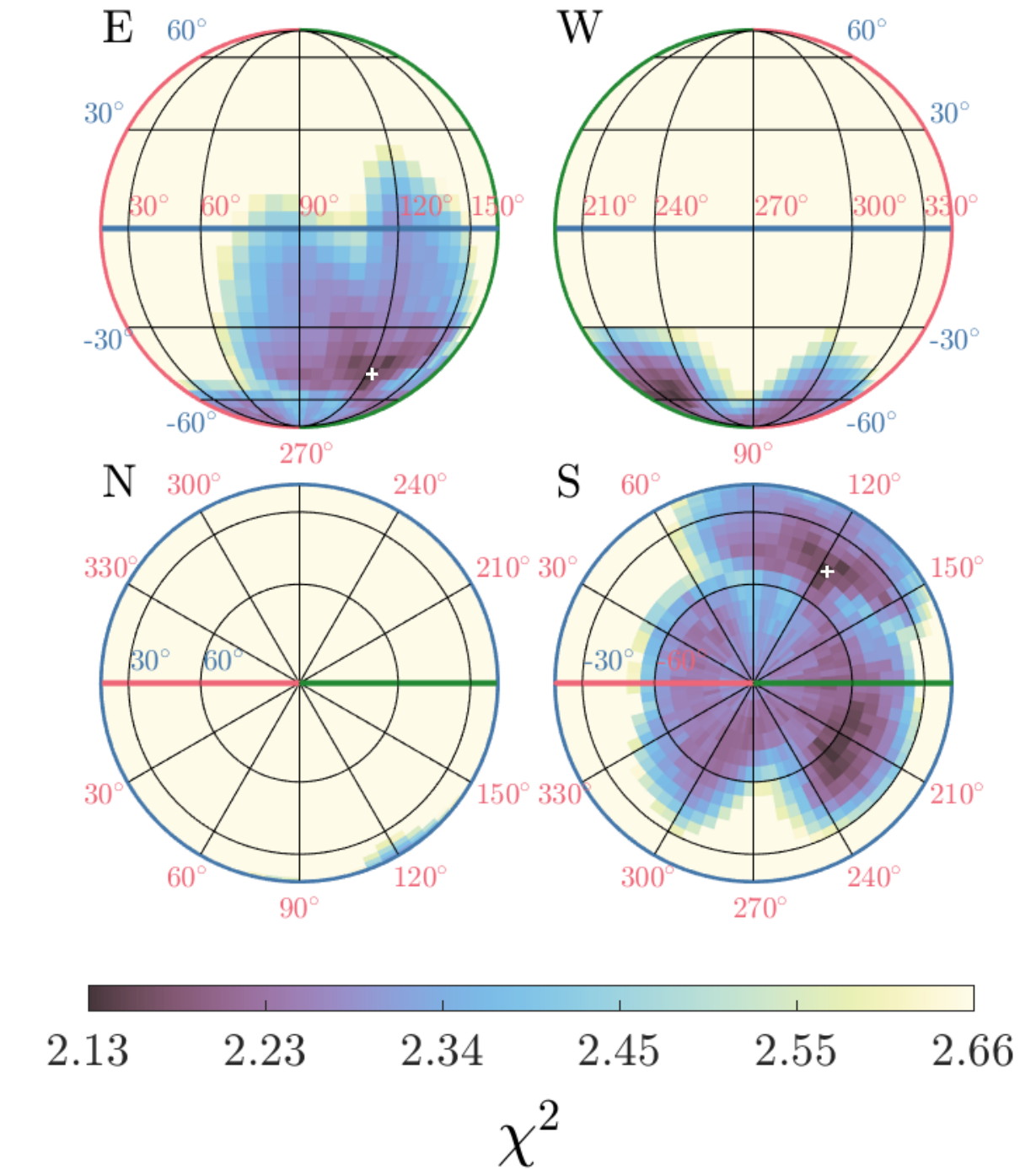}
    \caption{
    The distribution of $\chi^2$ values corresponding to likelihood of the nucleus rotation pole orientation, measured at $5\times5$ deg intervals over the entire ecliptic plane. These $\chi^2$ values have been projected onto a sphere of ecliptic longitude and latitude. The four spheres show the same solution from four viewing angles, along the cardinal direction labelled at the top left corner of each sphere. The darkest regions indicate the lowest values of $\chi^2$ where the pole is most likely oriented. The best-fit pole orientation is located at $(\lambda_E, \beta_E) =(120^\circ, -50^\circ)$ and is marked on the figure by a white cross in the E and S views. A second region of low $\chi^2$ is identifiable around $(\lambda_E, \beta_E) =(220^\circ, -60^\circ)$. }
    \label{fig:chiplane}
\end{figure}

To search for the optimal nucleus shape and rotation pole orientation, we created a coarse $5\times5$ deg grid of ecliptic coordinates covering the entire celestial sphere (longitude $0^\circ \leq \lambda_E < 360^\circ$ and latitude $-90^\circ\leq \beta_E \leq 90^\circ$). At each combination of pole coordinates, holding the sidereal period fixed, we ran \textsc{convexinv} with 50 iterations to find preliminary fits for each of the shape, spin state and phase function parameter $k_s$ (from Equation \ref{eq:pf}).
The overall distribution of pole solutions does not depend on $k_s$ - the value of this parameter affects only the detailed features of the shape model at each pole orientation. Therefore we fix $k_s$ at the best-fit value from the initial pole search, and re-run the full sky grid search with 500 iterations to identify the pole orientations to search with higher resolution.
The quality of the shape and spin state fit at each possible pole location are shown as a spherical projection of the $\chi^2$ plane in Fig. \ref{fig:chiplane}. The pole direction corresponding to the lowest value of $\chi^2$ at this stage lies at $(\lambda_E, \beta_E) = (120^\circ, -50^\circ)$. A second region of low $\chi^2$ values centred around $(\lambda_E,\beta_E)=(220^\circ,-60^\circ)$ is also identifiable from this distribution of pole solutions.

We refined the solution by performing a $2\times2$ deg grid search over a smaller region around this location, incorporating both areas of low $\chi^2$. At each grid point we fit both the shape and $k_s$ parameters with 500 iterations to ensure that $\chi^2$ was minimised. The best fit pole orientation was found to be at $(\lambda_E, \beta_E) = (118^\circ, -50^\circ)$. The uncertainty associated with these values was taken as the standard deviation in the $\lambda_E$ and $\beta_E$ values within 10 per cent of the $\chi^2$ minimum, giving final values $\lambda_E=(118^\circ\pm26^\circ)$ and $\beta_E=(-50^\circ\pm21^\circ)$. This solution implies an orbital obliquity of 167\textdegree\space corresponding to retrograde nucleus rotation. The optimised facet areas and normals were transformed into a convex polyhedron by the \textsc{minkowski} procedure included in the software package. This shape was converted to a polyhedron with triangular facets by the \textsc{standardtri} procedure, to create the convex model shown in Fig. \ref{fig:model}. The values of the best fit \textsc{convexinv} parameters and their uncertainties are given in column A of Table \ref{tab:expo}. 

This grid search method of identifying the shape and pole solution that minimises $\chi^2$ means that \textsc{convexinv} optimises an independent shape at each pole. We present the `most-likely' shape as that which minimises $\chi^2$, but note that statistically other shapes are only marginally less likely. To that end, we examine the shapes produced at pole orientations that result in slightly larger values of $\chi^2$. We find that for poles within a few degrees of $(118^\circ, -50^\circ)$, the shape is not appreciably altered from the reported best-fit shape. Around the second region of low $\chi^2$ values identified previously at $(\lambda_E,\beta_E)=(220^\circ,-60^\circ)$, the shape model found by the software is entirely unphysical. The model produced at this pole solution has a spin axis longer in length than its other dimensions, which implies an unstable rotation state. We therefore discard this as a possible solution, and continue this analysis using solely the best-fit model presented in Figure \ref{fig:model}.

\begin{figure*}
\centering
    \includegraphics[trim=0cm 0cm 0cm 0cm, clip=true, width=\linewidth]{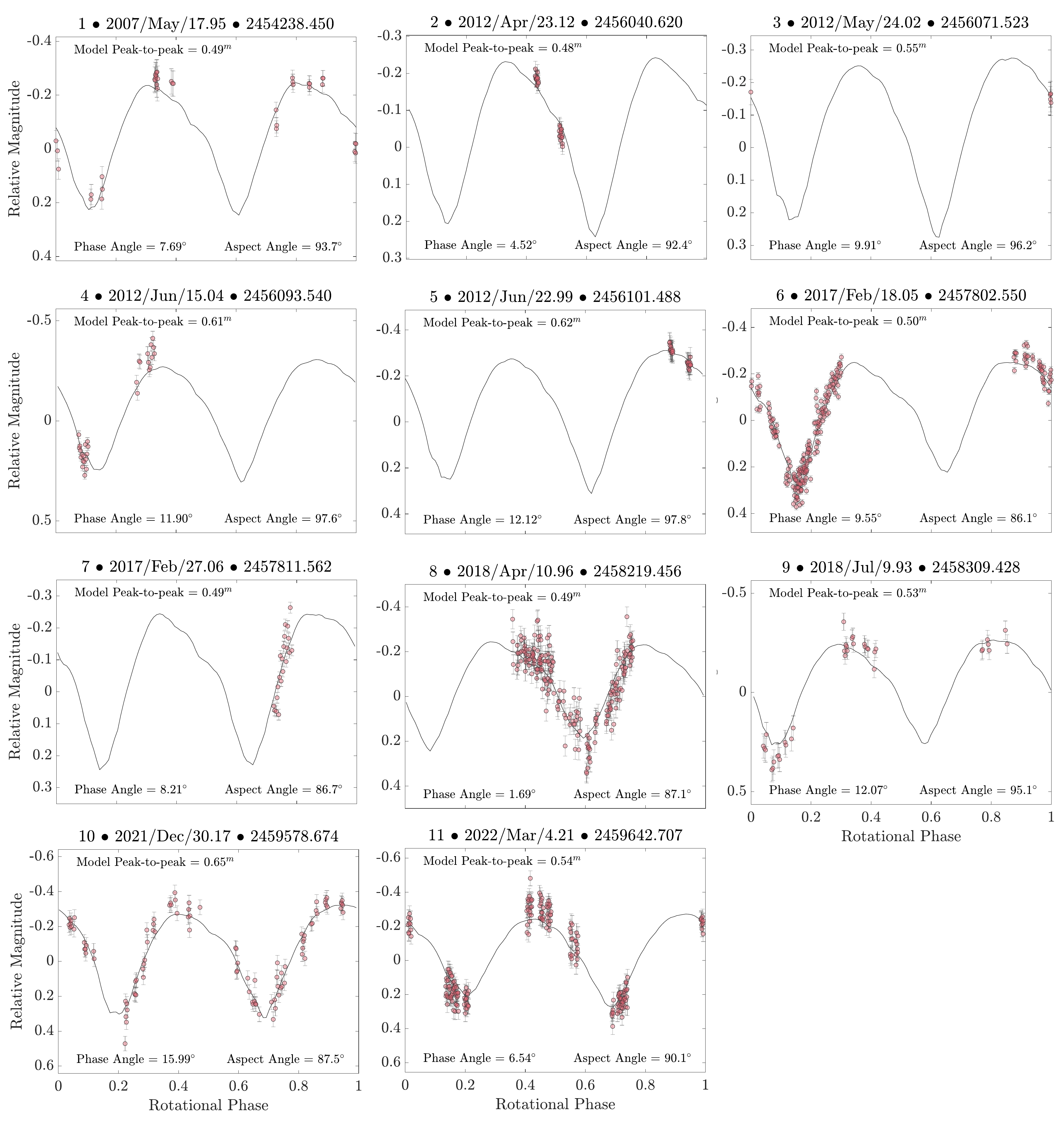}

\caption{Fits to the model lightcurve for all the observational epochs in the dataset. Lightcurves taken on successive nights (i.e. across nights when differences in viewing geometry are presumed negligible) have been combined and displayed under the date of the earliest night in the group. The observed lightcurve points have been scaled to have a mean of zero. The value given for the phase angle on each lightcurve is an average for all the observed points from the combined epochs.}
\label{fig:lcfits}
\end{figure*}

To examine how well this shape model matched the observed lightcurve points, we generated synthetic lightcurves for each observing epoch and compared these to the observed lightcurve points as shown in Fig. \ref{fig:lcfits}.

\subsubsection{Shape variability}
\label{infinitez}

\begin{figure*}
\centering
    \includegraphics[width=0.75\linewidth]{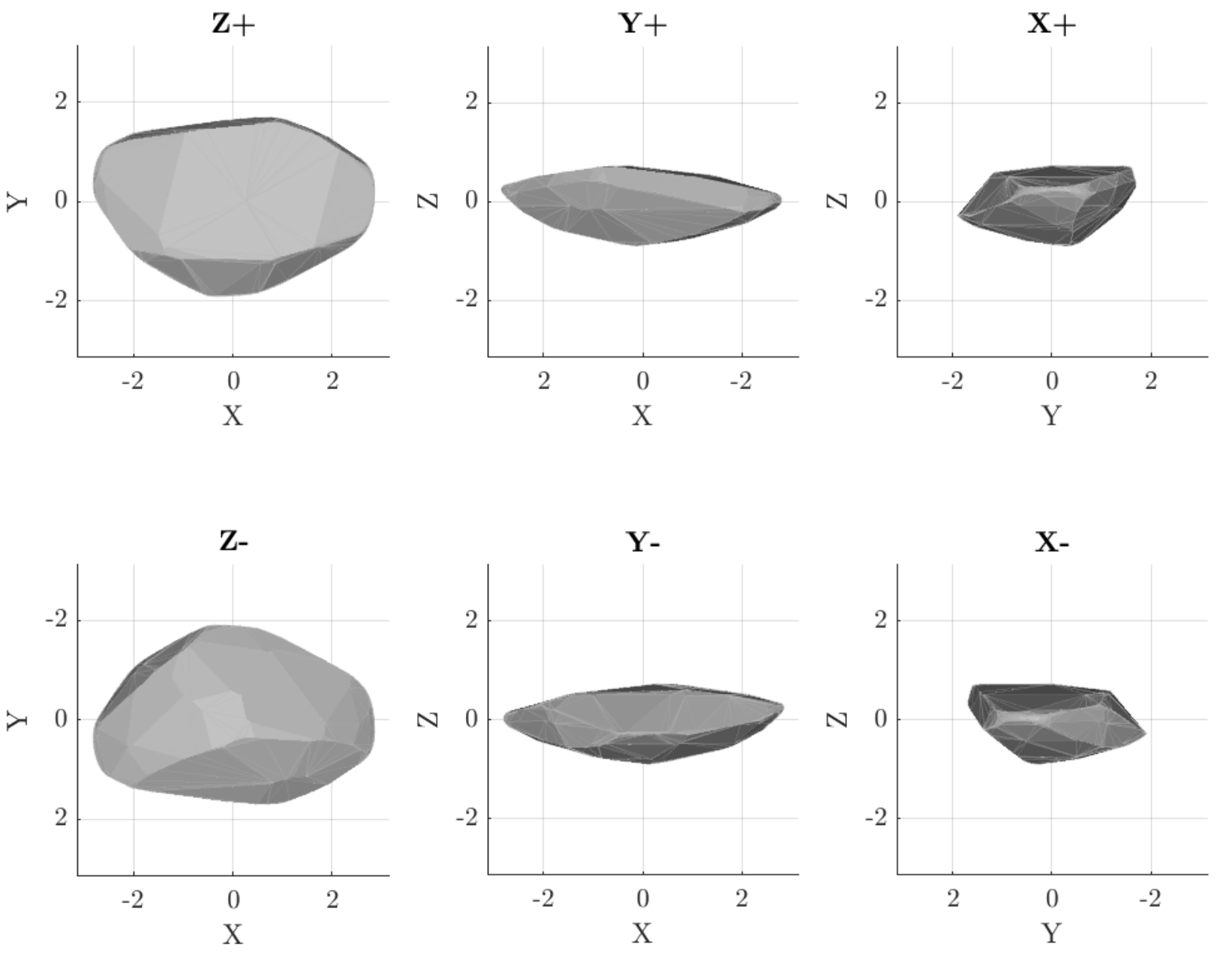}
\caption{Final shape model for 162P, viewed along three orthogonal directions. The model's rotation axis is aligned with the Z-axis of the plot, and the longest shape axis is aligned with the X-axis. The model results from the best fit sidereal period $P$ = 32.8638h, and rotation pole orientation ($\lambda_E, \beta_E) = (118^\circ, -50^\circ)$. The model is characterised by axis ratios $a$/$b$ = 1.56, $b$/$c$ = 2.33 and $a$/$c$ = 3.66 where $a,b,c$ are the three semi-axes of the equivalent-volume ellipsoid ($a>b>c$ in length).}
\label{fig:model}
\end{figure*}

\begin{figure*}
    \centering
    \includegraphics[width=\linewidth]{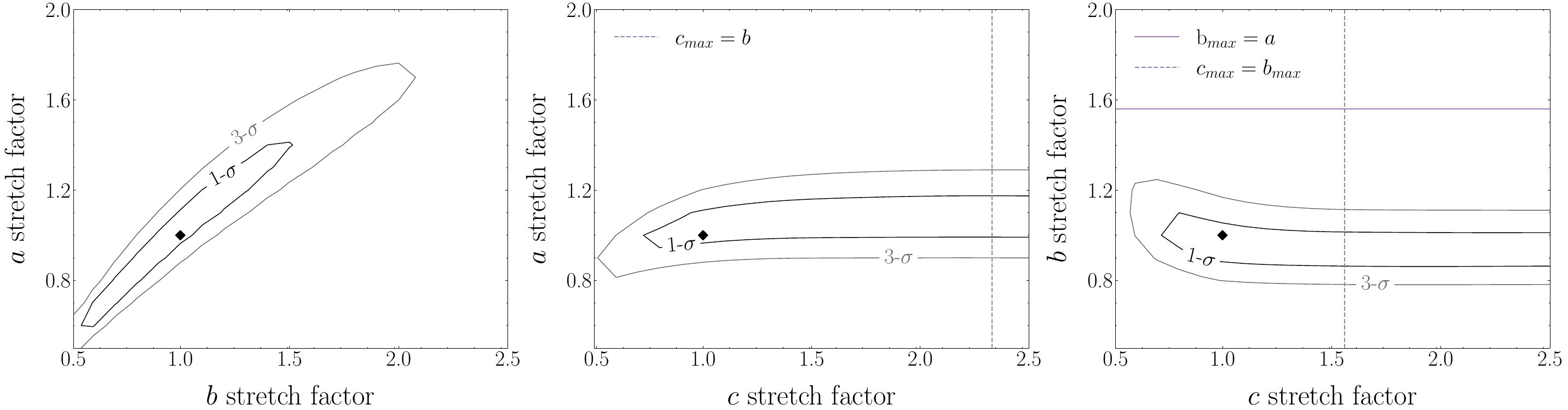}
    \caption{The 1-$\sigma$ and 3-$\sigma$ $\chi^2$ confidence levels that result when the best-fit shape for 162P is stretched along principal axes $a$ and $b$ (left), $a$ and $c$ (centre) and $b$ and $c$ (right). The third axis is held fixed in every case. The black point on each plot marks the location of stretch factor 1.0 for both axes. From the left figure, the possible variation in $a$ and $b$ is constrained at the 1-$\sigma$ level to 40 percent along $a$ and 45 percent along $b$, giving a possible range $1.4<a/b<2.0$. From the contours in the centre and right plots, it is clear that we can place a lower limit on the possible length of the $c$ axis, but the upper limit for the $\chi^2$ contours are unconstrained in both cases. The vertical dashed line in the centre plot demonstrates the physical upper limit $c=b$ for this model from rotational stability arguments. The solid horizontal line in the right plot shows the additional limit imposed by not allowing $b$ to exceed $a$ in length, and the vertical dashed line indicates how this limits the possible length of $c$.}
    \label{fig:stretch}
\end{figure*}

We employed a method similar to that of \cite{lowry2012nucleus} to explore to what extent we could vary the shape shown in Fig. \ref{fig:model} and still obtain a statistically-significant fit to the lightcurves. We created a grid of factors in range $0.5-2.5$ by which to stretch the lengths of the principal axes $a,b$ and $c$ of the shape's equivalent-volume ellipsoid. This enabled us to stretch along two shape axes at once while holding the length of the third axis fixed. We generated synthetic lightcurves for all resulting stretched models, and determined a $\chi^2$ value according to how well each stretched model lightcurve matched the observed lightcurve points (it should be noted that this value of $\chi^2$ was calculated as standard, and is not the same as the relative $\chi^2$ metric used by \textsc{convexinv}). We show the resulting $\chi^2$ fits as contours at 1-$\sigma$ and 3-$\sigma$ confidence intervals for the three combinations of stretched axes in Fig. \ref{fig:stretch}. The range of possible axes ratios that $a/b$ can have while providing a statistically-significant fit to the lightcurve points at the 1-$\sigma$ level is $1.4<a/b<2.0$. However, we are not able to constrain an upper limit for the $c$-axis in this way i.e. the length of the model's rotation axis. 
This is likely a direct result of the limited range of aspect angles at which the nucleus was observed - according to the best-fit pole orientation, the largest and smallest aspect angles at which 162P was observed vary by only 12\textdegree.

 We instead consider the rotational stability of the equivalent-volume ellipsoid that defines the axes $a, b, c$ along which the model is stretched. Since 162P is observed to be in a stable rotation state in every epoch, we assume that its axis of rotation, $c$, is the shortest in length. Should $c$ be stretched to be longer than $b$, then the model would no longer rotate about the axis with the largest moment of inertia, leading to unstable rotation. When $b$ is fixed, we can therefore constrain the maximum stretch factor applied to $c$ to be that which makes it equal in length to $b$. Imposing this upper limit means that the axis ratio $a/c$ can range from 1.5 to 5.5 and still provide a statistically significant fit to the lightcurve points at the 1-$\sigma$ level. For $b/c$, we impose that $b$ cannot exceed fixed $a$ in length, and that $c$ cannot exceed $b$. The resulting possible values for $b/c$ range from $1.5$ to $4.0$. The limits imposed from these stability-based arguments are shown as straight lines in the centre and right plots of Fig. \ref{fig:stretch}.

\subsubsection{Phase function considerations}
\label{testpf}

As described previously, we chose to force \textsc{convexinv} to fit a shape from the lightcurves using a purely linear phase function. The form of Equation \ref{eq:pf} emulates empirically the typical behaviour of the surfaces of Solar System objects: increasing linearly in brightness with decreasing phase angle $\alpha$, and surging in brightness exponentially around $\alpha\sim0^\circ$ in what is known as the opposition effect (OE). The phase functions of JFC nuclei acquired using ground-based observations have not yet revealed any evidence for a cometary OE. In every case, a purely-linear phase slope $\beta$ provides a suitable fit to the lightcurve points \cite[reviewed in][]{snodgrass2011size, 2017kokotanekova}. 

Comet nucleus observations require negligible activity and therefore generally occur at large heliocentric distances, resulting in relatively small solar phase angles. However, nucleus observations at phase angles close to zero are extremely rare because they require an alignment of orbital nodes with the opposition direction. The 2018 lightcurves (ID 13-15) captured the nucleus of 162P close to phase angle zero for the first time, providing the opportunity to detect an OE if it occurred. To explore this possibility, we performed the shape modelling procedure twice: once exactly as described with a linear phase function; and the second time allowing \textsc{convexinv} to also fit the parameters $a_s$ and $d_s$ that characterise the exponential component of Equation \ref{eq:pf}, as well as $k_s$ in the phase function. The best-fitting shape model from this test was practically identical to the original shape model, as summarised in Table \ref{tab:expo}.

\begin{table}
 \centering
 \caption{A comparison of the shape model properties resulting from shape optimisation with a linear phase function (A) and with a linear-exponential phase function (B). $a_s,d_s$ and $k_s$ are the best-fit parameters of the phase function described in Equation \ref{eq:pf}. The values $a/b$ and $b/c$ are the axis ratios calculated for an ellipsoid with the same volume as the shape model. The properties of the shape fit with a linear-exponential phase function are virtually identical to the shape produced with the linear phase function.}
 \label{tab:expo}
 \begin{tabular*}{0.75\linewidth}{ccc}
  \hline
  Shape property & A & B \\
  \hline
 $P$ [h] & $32.864\pm0.001$ & $32.864\pm0.002$ \\
 $\lambda_E$ [\textdegree]& $118\pm26$ & $118\pm23$ \\
 $\beta_E$ [\textdegree] & $-50\pm21$  & $-50\pm28$ \\ 
 $a_s$ & - & 0.04 \\ 
 $d_s$ & - & 0.01 \\ 
 $k_s$ & -1.47 & -1.48 \\ 
 $a/b$ & 1.56 & 1.58 \\ 
 $b/c$ & 2.33 & 2.24 \\ 
   \hline
 \end{tabular*}
\end{table}

\begin{figure}
    \centering
    \includegraphics[width=\linewidth]{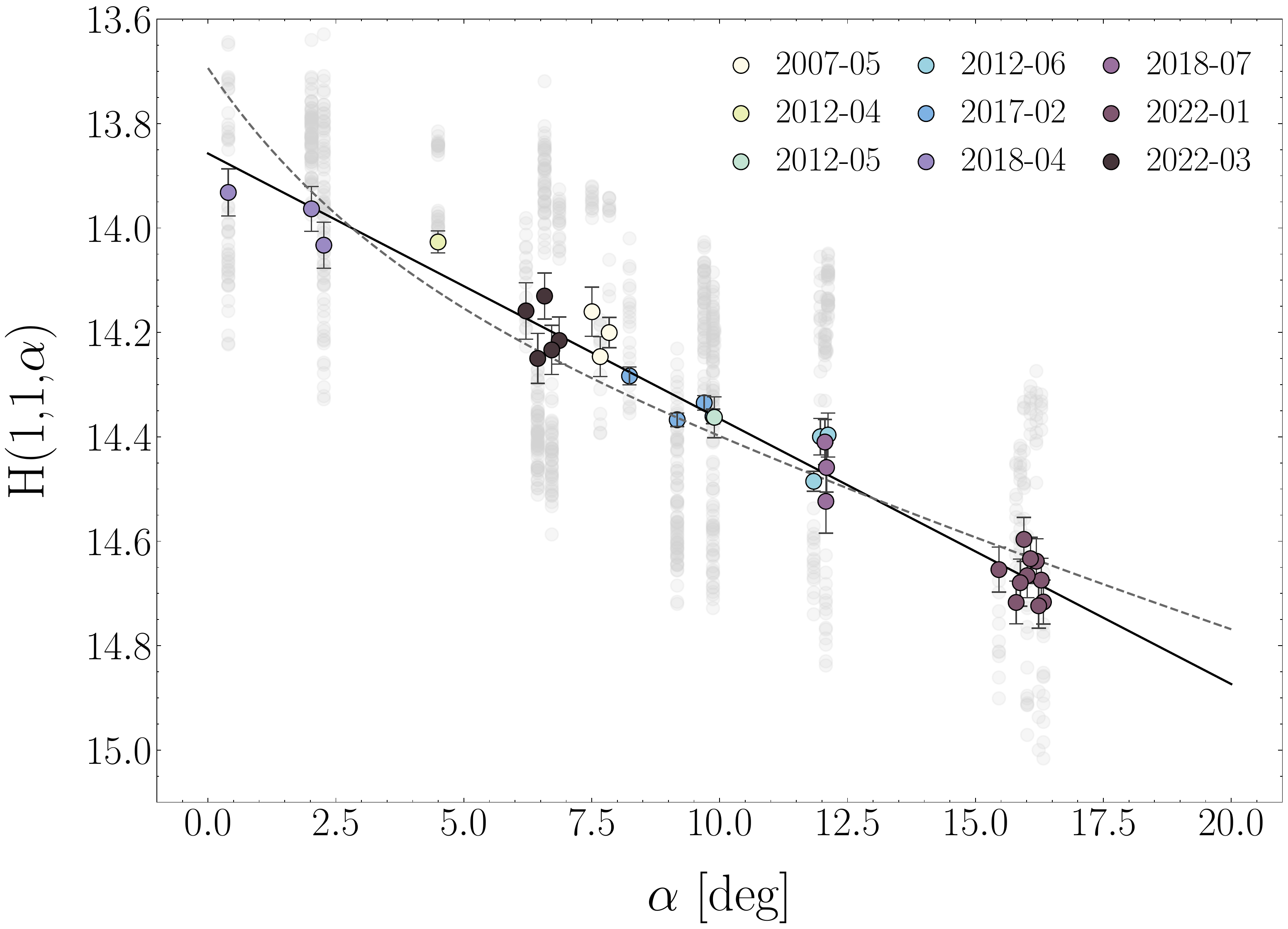}
    \caption{The phase function for 162P. Every lightcurve (ID 1-33) has been corrected for rotational effects using a synthetic lightcurve generated from the shape model at that observing geometry, to obtain a rotationally-averaged magnitude for each night. The black solid line depicts the best linear fit to these data points, with a slope value of $\beta=0.051\pm0.002$ mag deg$^{-1}$ and intercept $H_r(1,1,0)=13.857\pm0.020$ . The dashed line shows the best $H,G$ fit to the points, which provides a poor fit to the data, particularly at low phase angles. The best-fit values for the $H,G$ function are $H=13.694\pm0.022$ and $G=0.094\pm0.025$. The original lightcurve magnitudes (with no rotational correction) are shown in grey.}
    \label{fig:phasecurve}
\end{figure}

With the addition of the 2018, 2021 and 2022 lightcurves, we expanded 162P's observed phase angle coverage to $0.4^\circ<\alpha<16.4^\circ$ (see Table \ref{tab:obs}). Using the shape model, it is possible to remove the effects of rotation from the lightcurve points, allowing the phase function to be determined with magnitudes at a consistent equatorial geometry in what is known as a reference phase curve, defined by \cite{kaasa2}. The reference phase curve for 162P is shown in Fig. \ref{fig:phasecurve}. We fit two phase functions to the points: a linear slope and an $H,G$ function \citep{bowellhg}. The $H,G$ model provides a poor fit to the data points, particularly at small phase angles, while the linear function is in excellent agreement with the rotationally-corrected lightcurve points, indicating that 162P does not show evidence for an opposition surge in this dataset. From the linear fit, the phase function is characterised by a slope value of $\beta=0.051\pm0.002$ mag deg$^{-1}$ and intercept $H_r(1,1,0)=13.857\pm0.020$. 
We use this value for $H_r$ to calculate 162P's $r$-band geometric albedo ($p_r$) using Equation \ref{eq:albedo}.

\begin{equation}
    \label{eq:albedo}
    p_{r,PS1} = (k^2 / R^2) \times 10^{0.4(m_\odot -  H_r)}
\end{equation}

\noindent Here, $k = 1.496\times10^8$ km is the conversion factor between astronomical units and kilometres, m$_\odot$ is the $r$-band PS1 magnitude of the Sun (-26.91 mag) and  $R=7.03^{+0.47}_{-0.48}$ km is the effective radius of the comet nucleus. It should be noted that this value for $R$ was obtained using thermal IR measurements \citep{fernandez2013thermal}. Assuming that the nucleus size has remained unchanged, this yields a geometric albedo $p_r=0.022 \pm 0.003$. To compare this with the existing literature values, we convert $H_r$ to $H_R$ and $H_V$ using the conversions described in \cite{tonry2012pan}, and derive values for $p_R=0.023\pm0.003$ and $p_V=0.021\pm0.002$. The value for $p_R$ is consistent with the values obtained by \cite{2017kokotanekova, 2018kokotanekova} within the uncertainties, implying that 162P remains one of the darkest surfaces of all studied JFCs. 

\section{Discussion}
\label{disc}

\subsection{Nucleus properties from shape modelling}

\subsubsection{Shape}

 The best-fitting shape model for 162P shown in Fig. \ref{fig:model} has an elongated axis ratio $a/b=1.56$, and appears relatively flat when viewed along the rotation pole. The stretching analysis performed in Section \ref{infinitez} revealed that the axis ratio $a/b$ can range from $1.4-2.0$ and still provide a suitable fit to the lightcurve points at the 1-$\sigma$ level. Typically for comets the lightcurve amplitude $\Delta m$ is used to place a lower limit on the axis ratio, since the orientation of the pole with respect to the observer is generally not known. For all known JFCs, the mean value for this lower limit $a/b=1.5$ \citep{2017kokotanekova, lamy2004sizes}. 
 
 The axis ratios of the spacecraft-visited JFCs have been estimated more precisely from measurements of their shape dimensions. Using the literature values, we determine that they have a mean value $a/b=2.0$ \citep{buratti2004, thomas20139p, 2004_81p, thomas2013shape, jorda2016global}. When considering solely the bilobed JFCs (19P, 67P and 103P), this mean value becomes 2.4, and for 9P and 81P which are not bilobed, the average $a/b$ is 1.3. To determine these values we have used the ratio of the longest measured axis to the second-longest, contrary to the works stated above which quote $a/b$ for spacecraft targets as the ratio of longest to shortest axis length. To directly compare these values to the shape derived in this work, we calculate the relative lengths of the axis of the shape model to be: $a/b=1.6$; $b/c=2.2$; and $a/c=3.5$. The axis ratio $a/b$ indicates that the nucleus of 162P is more elongated than the non-bilobed JFCs, but not enough to match the average properties of the currently-known bilobed JFCs. It should be noted that we have not included 1P/Halley in this brief analysis. This is due in part to the fact that it is, by definition, not a JFC, and also to the contentious nature of describing it as bilobed. Adding the axis ratio for 1P, $a/b=1.96$, does not change the mean axis ratio for the spacecraft-visited comets, and yields $a/b=2.3$ for the mean elongation of the bilobed comets. 

For the five spacecraft-observed JFCs, the mean ratio of the longest-to-shortest axes $a/c$ is 2.2. This ratio becomes 2.7 when considering only the bilobed JFCs, and 1.5 for the non-bilobed. The value derived for 162P is 3.5, considerably higher than both of these estimates, and is similar to the dogbone-shaped JFC 103P. It is possible that the nucleus of 162P is long and relatively flattened in shape - objects in such configurations are known to exist in the Solar System, for example Arrokoth. A recent study of 3552 Don Quixote, a near-Earth object suspected to be of cometary origin, also resulted in a elongated and relatively flattened shape model using CLI \citep{Mommert_2020}. 

The XY-plane of the 162P shape model is dominated by large regions of flat facets, which may be masking large-scale concave surface features that cannot be recreated by the lightcurve inversion procedure. The combination of these flat facet regions and the model axis ratios present tentative evidence for a bilobed shape: for example, the flat regions may be masking a slim neck connecting two distinct lobes. This interpretation should be treated with caution, as it is well-established that inferring concavities from lightcurves alone is not conclusive \citep{contactbrickary}. Moreover, as demonstrated in Section \ref{infinitez}, we were unable to constrain a maximum length for $c$ (model $Z$-axis) solely from fits to the observed lightcurve points at the 1-$\sigma$ level. It is therefore possible that the $Z$-axis has been underestimated somewhat by the convex inversion procedure, given the limited variation in viewing geometry covered by the lightcurves. While \textsc{convexinv} imposes no constraints on the possible shape dimensions and identified this model as the shape that best fits the input lightcurves when accounting for all variables, if the $Z$-axis (i.e. rotation axis) was continually oriented away from Earth then the true extent of this axis is almost impossible to quantify. However, the minimum value for both $a/c$ and $b/c$ was constrained to be 1.5. We therefore suggest that the nucleus of 162P is more elongated than the non-bilobed comets for which we have detailed shape information, and that its longest axis is at least 1.5 times greater in length than its axis of rotation.

\subsubsection{Phase function}

We have used two different methods to measure the phase function of 162P in this work. We derived a value for $\beta=0.0468\pm0.0001$ mag deg$^{-1}$ using a Monte Carlo method on the entire lightcurve (without the removal of rotational effects). The uncertainty produced by this method is deceptively small, and accounts for the photometric and absolute-calibration uncertainties only. By correcting the lightcurves for the effects of rotation with the convex shape model, and fitting a line to the rotationally-averaged lightcurve points as a function of phase angle, we obtained a phase function value $\beta=0.051\pm0.002$ mag deg$^{-1}$. %
This value of $\beta$, derived using the shape model, is slightly steeper than the value obtained without correcting the lightcurve points for rotational effects.

The newly derived 162P phase coefficient is larger than the $\beta=0.039\pm0.002$ mag deg$^{-1}$ determined by \citet{2017kokotanekova} using the 2007-2017 datasets (IDs 1-12). The updated value of the phase function slope is derived for a broader range of phase angles (expanding the previous range of $4^\circ-12^\circ$ to $0.4^\circ-16^\circ$). Moreover, modeling the nucleus shape and correcting the photometry using the shape model accounts for the lightcurve shape effects which was not possible in \cite{2017kokotanekova, 2018kokotanekova}. We therefore adopt the new PS1 $r$-band phase coefficient $\beta=0.051\pm0.002$ mag deg$^{-1}$ and the corresponding absolute magnitude $H_r=13.857\pm0.020$, and conclude that the phase function of 162P is steeper than previously determined.

With a minimum phase angle of $\alpha\sim0.4^\circ$, and a total of four observing epochs at phase angles $\alpha<5^\circ$, the phase function of 162P is unique among other comet nuclei observed from the ground. Only two other JFCs have published nucleus lightcurves at phase angles less than $1^\circ$: 28P/Neujmin 1 \citep{2001delahodde} and 137P/Shoemaker-Levy 2 \citep{2017kokotanekova}. 
The geometric albedos of these objects are $0.03\pm0.01$ \citep{jewittmeech1988, campins1987} and $0.034\pm0.006$ \citep{2017kokotanekova} correspondingly. Like 162P, neither of these objects display any evidence for an opposition effect. To date, the only comet nucleus with a clearly detected OE is 67P \citep{fornasier2015}. However, this OE was {\em not} detected from the ground - rather, it was observed in situ by the Rosetta spacecraft from the disk-integrated flux of the resolved nucleus. The geometric albedo of 67P is $p_R=0.065\pm0.002$ \citep{fornasier2015}, which is more than double the measured albedos of 162P, 28P and 137P. It is therefore not entirely unexpected that 162P did not demonstrate evidence for an opposition effect in the phase angle range covered. Its surface is substantially darker than 67P, and darker still than 28P and 137P, both of which exhibited no OE. Moreover, other minor planet populations with low albedos such as Centaurs \citep{belskaya2000opposition} and some Jupiter Trojans \citep{shevchenko2012} are also known to display very narrow opposition effects with amplitudes less than 0.2 mag below phase angles $\sim$$0.1 - 0.2^\circ$.


This work's finding that the phase function of 162P is steeper than previously estimated, while its albedo remains very small, brings into question whether 162P is in agreement with the potential relationship between geometric albedo and phase slope identified in \cite{2018kokotanekova}. That work compiled a database of 14 JFCs with well-constrained albedos and phase coefficients, and discovered a possible trend of increasing $\beta$ with increasing albedo. This behavior is opposite to the correlation between the linear phase coefficient and geometric albedo of asteroids \citep{belskaya2000opposition}, and was interpreted as a possible evolutionary trend for comet nuclei in which the lowest-albedo comets have the most evolved surfaces. The new, steeper phase coefficient value of 162P places it closer to the asteroid correlation than the one found for JFCs. Since this is the first low-albedo comet phase function derived after accounting for the nucleus shape model, this finding challenges the possible correlation between comet phase functions and albedos.
In addition, the phase function of 28P was initially reported by \citet{2001delahodde} as $\beta=0.025\pm0.006$ mag deg$^{-1}$ over a phase angle range of $\sim$$0-15^\circ$. \citet{schleicherDPS} more recently determined a significantly steeper value for its phase function, $\beta\sim0.05$ mag deg$^{-1}$. This result makes 28P a potential second outlier for the phase function-albedo correlation hypothesis, further contesting this proposed evolutionary trend.

Alternatively, if we continue to assume that the other comets in the sample have well-constrained phase functions, we must explain what makes 162P (and potentially 28P) the exception with a steep phase function and very small geometric albedo. One possible explanation for this discrepancy might come from the orbital analysis of the JFCs in the near-Earth space by \citet{fernandez2015}. In that work, the orbital integration of 162P revealed that it may originate from the outer Main Belt rather than from the Scattered Disk population currently believed to be the main source of JFCs. It is therefore constructive to compare the surface properties of 162P to those of very dark asteroids. However, very few low-albedo asteroids have well-constrained phase functions. Two notable examples are asteroids (101955) Bennu and (162173) Ryugu, studied in-situ by OSIRIS-REx and Hayabusa2 respectively. They have geometric albedos $p_V=0.044\pm0.002$ \citep{Dellagiustina2019} and $p_V=0.040\pm0.005$ \citep{tatsumi2020}, larger than that estimated for 162P, and both have a small opposition surge detectable at $\sim$ $0 - 7$\textdegree\space \citep{2019hergenrother, tatsumi2020}. To our knowledge, the only asteroid with comparable geometric albedo to 162P is the Jupiter Trojan (1173) Anchises with geometric albedo $p_V = \mathrm{0.027^{+0.006}_{-0.007}}$ \citep{2012horner}. Interestingly, the phase function of Anchises at phase angles $0.3-2^\circ$ is remarkably shallow with $\beta=0.023\pm0.008$ mag deg$^{-1}$. 

Such a small sample of objects with both geometric albedo $<$0.05 and a well-constrained phase function offers limited possibilities to understand the unusual properties of 162P and 28P. However, future targeted campaigns or photometric data from surveys such as Rubin Observatory's Legacy Survey of Space and Time (LSST) provide the opportunity to significantly increase the number of objects with reliable phase functions over large phase angle ranges, and could enable us to build better statistics in order to test the ideas proposed to explain why 162P's phase function is steeper than those of other, similarly dark JFCs.

\subsection{Challenges associated with cometary convex inversion}

Comets are generally active within $\sim$3 au of the Sun, which is also where they are closest to Earth and most observable. As such, in most cases we cannot get reliable ground-based nucleus photometry over these parts of their orbits. We are therefore limited to observing when they are at heliocentric distances $\gtrsim 3$ au, yet still bright enough for sufficient signal-to-noise. Coupling this with the low orbital inclinations of the JFCs resulting in a limited range of body-centric latitudes visible to the observer, it is challenging to observe comet nuclei at a wide range of viewing geometries.%

The range of observing geometries obtained for 162P is sizeable for a short period comet nucleus, and is comparable to that of asteroids that have been previously modelled by convex inversion. This is likely due to 162P's large effective radius, which means that the comet remains sufficiently bright around aphelion. In addition, its low levels of activity mean that the nucleus signal is not typically contaminated by dust coma at heliocentric distance $\sim$3 au, making it an excellent candidate for targeted observations with limited telescope time available. The observer-centred ecliptic longitude and latitude of 162P in our dataset span a range of $\sim$$48^\circ$ and $\sim$$22^\circ$ respectively. In comparison, \citet{lowry2012nucleus} obtained observations of the nucleus of 67P at observer-centred ecliptic longitude and latitudes over a range of $100^\circ$ and $15^\circ$ respectively to create its convex model. Their observations only varied in aspect angle by $17^\circ$ (between $53^\circ-70^\circ$) before the authors included images from HST at an aspect angle of $100^\circ$. For 162P, the best-fitting pole solution was oriented towards $(\lambda_E,\beta_E) = (118^\circ, -50^\circ)$ (corresponding to R.A.$=290$\textdegree, decl.$=-28$\textdegree) implying that the lightcurves spanned a $12^\circ$ range of aspect angles from $86^\circ-98^\circ$.

Rotation pole information is rare for low activity comets, because the most common means of determining pole orientation involves tracking or modelling the temporal evolution of morphological structures in the coma. The statistics for cometary pole distributions are therefore extremely limited. It is expected that the distribution of pole orientations for the population to be somewhat random, given that the torques exerted by the sublimation of surface volatiles are capable of dramatically altering the nucleus spin state \citep{samarasinha2004rotation}. Comparing the pole orientation derived for 162P to the limited existing orientations known for other comets is therefore unlikely to offer any particular insights into the properties of the population. 

\begin{figure}
    \centering
    \includegraphics[width=\linewidth]{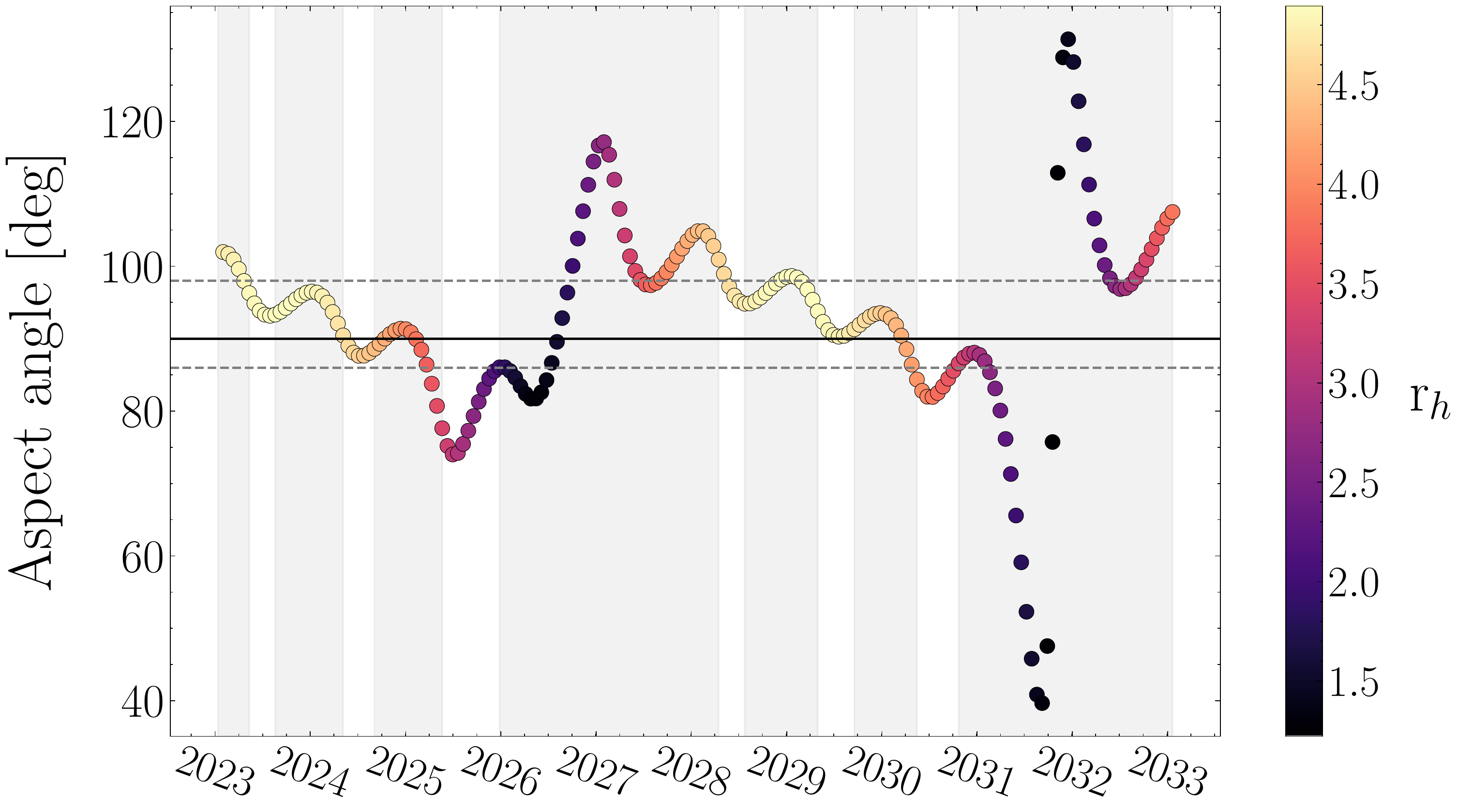}
    \caption{Variation in expected aspect angle of comet 162P over the approximate LSST operation period (early 2023-2033), extrapolated based on its present-day orbital elements and best-fit pole solution. The solid black line indicates the average aspect angle covered by the existing 2007-2022 observations, and the dashed lines show the maximum and minimum aspect angles covered by this dataset. The colour of each point corresponds to the heliocentric distance $r_h$ of the comet at that time, with the darkest coloured points being closest to the Sun. Grey background indicates where observations of 162P are not obtainable by LSST.}
    \label{fig:aspect}
\end{figure}

Placing constraints on the pole orientation allows us to predict future geometries that may be available to further enhance the model and refine the z-axis. Prior to this work, it was possible to infer that 162P had been observed at a limited range of aspect angles due to the similar $\Delta m$ values of lightcurves obtained at different epochs. The upcoming LSST has the potential to extend the range of observing geometries for 162P and many other JFCs. To illustrate this, we used the pole solution to determine the aspect angle for 162P based on its current orbit, at 20-day intervals over a time frame which coincides approximately with LSST. The expected variation in aspect angle is shown in Fig. \ref{fig:aspect} as well as the comet’s heliocentric distance at each timestamp.  It should be noted that these points do not account for whether or not 162P will actually be observable by LSST - those that are not observable are plotted on a grey background. The figure shows that the nucleus exhibits a much greater range of aspect angles than those covered by our dataset, around 90$^\circ$. However, the extremes in aspect range are all at low heliocentric distance $<2.5$ au, where observations are more likely to be affected by activity. This demonstrates the difficulty of implementing CLI on comet nuclei; unlike asteroids which are best observed close to Earth during which time they move quickly along the sky and offer a wide range of viewing angles, comets must be observed further out, considerably limiting the available viewing geometries. 

Nevertheless, LSST will produce an abundance of calibrated, temporally-sparse photometry for many comets, resulting in a much greater range of aspect angles than is realistic using present methods of targeted observations such as the lightcurves presented in this work. It has been shown to be possible to create reliable shape models of asteroids using sparse photometry from survey data combined with densely-sampled lightcurves \citep{durech2009sparse}. With a greater range of observing geometries available for a large number of known comets, we expect that it will prove possible to constrain the shapes and pole orientations of many more of these objects from ground-based observations alone.

\section{Summary}

This works presents photometric lightcurves of the nucleus of Jupiter-family comet 162P/Siding Spring obtained over 33 epochs between 2007-2022 when no cometary activity was detected. The dataset was analysed with the aim to derive a convex shape model of the comet's nucleus.

\begin{itemize}
    \item We collected new lightcurves for comet 162P in 2018 and 2021/2022, including in April 2018 when the comet was close to 0\textdegree\space phase angle. With the addition of these observations, the total range of phase angles at which the nucleus has been observed was extended from $4^\circ-12^\circ$ to $0.4^\circ-16.3^\circ$.
    \item Using photometry obtained in March 2022, we determined a $(g-r)$ colour estimate for the nucleus (in Pan-STARRS 1 filters) of $0.48\pm0.04$ mag. This value is consistent with previous estimates of the nucleus colour.
    \item We used this colour and the $r$-band magnitudes of background stars in the Pan-STARRS 1 catalogue to absolutely-calibrate the comet lightcurves, and update the calibration of existing photometry of the comet obtained in 2007-2017.
    \item We used convex lightcurve inversion on the calibrated lightcurves to fit a shape model, pole orientation and sidereal period for the nucleus. Using a linear phase function to model the lightcurves, we obtain a shape with axis ratios $a/b=1.56$ and $b/c=2.33$ and pole orientation $(\lambda_E, \beta_E)=(118^\circ\pm26^\circ, -50^\circ\pm21^\circ)$.
    \item To examine the extent to which this shape can vary while still fitting the lightcurves with statistical significance, we applied combinations of stretching factors to its principal axes, two at a time. We found that the $a/b$ axis ratio can vary between 1.4 and 2.0 and still fit the lightcurves at the 1-$\sigma$ level. We found that we could not place a statistical limit on the upper length of the rotation axis from the lightcurves alone, due to the limited range of observing geometries covered by the lightcurves. Using rotational stability arguments, the range of possible values for $b/c$ was constrained to $1.5<b/c<4.0$.
    \item With the best-fit shape model we corrected the lightcurves for the effects of rotation, and fitted both a linear and $H,G$ phase function to the resulting distance-corrected magnitudes. The $H,G$ model provided a poor fit to the data, while the linear phase function $\beta=0.051\pm0.002$ mag deg$^{-1}$ matched the datapoints extremely well at all phase angles. We concluded that 162P did not display a detectable opposition surge in the 2018 lightcurves, and discussed the implications that a steep phase slope for an object with such a dark surface has on current hypotheses for comet surface evolution.
    \item We suggested that 162P may be another example of a bilobed JFC based on the tentative evidence provided by the large planar regions in the model $XY$-plane, and its elongation. 
    \item We used the best-fit pole solution from the convex lightcurve inversion procedure to predict changes in the comet viewing geometry throughout the decade in which LSST will be operational which would allow us to further refine the shape. The aspect angle varies most extensively at heliocentric distances closest to the Sun, highlighting the challenges of obtaining reliable observations of comet nuclei at a range of viewing geometries.

\end{itemize} 

\section*{Acknowledgements}

The authors would like to thank the referee, David Schleicher, for insightful and helpful comments on this manuscript. We also thank Samuel Jackson for the useful discussions.

This work was supported by the UK Science and Technology Facilities Council.
This work was also facilitated by support from the International Space Science Institute in the framework of International Team 504 "The Life Cycle of Comets".
RK acknowledges support by ESO through the ESO Fellowship. This work was supported in part by ESO's SSDF 21/22 (Student) Garching funding program.

The new lightcurves presented were based on observations made with the Isaac Newton Telescope (UK PATT programmes I/2018A/07 and I/2022A/07), the Liverpool Telescope (programme XPL21B13) and the ESO New Technology Telescope (programme 0101.C-0709(A)). 
Previously reported lightcurves were based on observations at the European Southern Observatory under ESO programmes 089.C-0372(A) and 089.C-0372(B), the WHT and INT under UK PATT programmes W\slash 2007A\slash20 and I\slash2017A\slash05, and the 2-m telescope at Rozhen Observatory, Bulgaria.

The Pan-STARRS 1 Surveys (PS1) and the PS1 public science archive have been made possible through contributions by the Institute for Astronomy, the University of Hawaii, the Pan-STARRS Project Office, the Max-Planck Society and its participating institutes, the Max Planck Institute for Astronomy, Heidelberg and the Max Planck Institute for Extraterrestrial Physics, Garching, The Johns Hopkins University, Durham University, the University of Edinburgh, the Queen's University Belfast, the Harvard-Smithsonian Center for Astrophysics, the Las Cumbres Observatory Global Telescope Network Incorporated, the National Central University of Taiwan, the Space Telescope Science Institute, the National Aeronautics and Space Administration under Grant No. NNX08AR22G issued through the Planetary Science Division of the NASA Science Mission Directorate, the National Science Foundation Grant No. AST-1238877, the University of Maryland, Eotvos Lorand University (ELTE), the Los Alamos National Laboratory, and the Gordon and Betty Moore Foundation.

\section*{Data Availability}

The lightcurves used in this work can be accessed in an online Table A1 available at the CDS via anonymous ftp to cdsarc.u-strasbg.fr (130.79.128.5) or via http://cdsarc.u-strasbg.fr/viz-bin/cat/J/MNRAS. The shape model derived here will be shared on reasonable request to the corresponding author.


\bibliographystyle{mnras}
\bibliography{references}{}

\begin{thebibliography}{}
\makeatletter
\relax
\def\mn@urlcharsother{\let\do\@makeother \do\$\do\&\do\#\do\^\do\_\do\%\do\~}
\def\mn@doi{\begingroup\mn@urlcharsother \@ifnextchar [ {\mn@doi@}
  {\mn@doi@[]}}
\def\mn@doi@[#1]#2{\def\@tempa{#1}\ifx\@tempa\@empty \href
  {http://dx.doi.org/#2} {doi:#2}\else \href {http://dx.doi.org/#2} {#1}\fi
  \endgroup}
\def\mn@eprint#1#2{\mn@eprint@#1:#2::\@nil}
\def\mn@eprint@arXiv#1{\href {http://arxiv.org/abs/#1} {{\tt arXiv:#1}}}
\def\mn@eprint@dblp#1{\href {http://dblp.uni-trier.de/rec/bibtex/#1.xml}
  {dblp:#1}}
\def\mn@eprint@#1:#2:#3:#4\@nil{\def\@tempa {#1}\def\@tempb {#2}\def\@tempc
  {#3}\ifx \@tempc \@empty \let \@tempc \@tempb \let \@tempb \@tempa \fi \ifx
  \@tempb \@empty \def\@tempb {arXiv}\fi \@ifundefined
  {mn@eprint@\@tempb}{\@tempb:\@tempc}{\expandafter \expandafter \csname
  mn@eprint@\@tempb\endcsname \expandafter{\@tempc}}}

\bibitem[\protect\citeauthoryear{Belskaya \& Shevchenko}{Belskaya \&
  Shevchenko}{2000}]{belskaya2000opposition}
Belskaya I.,  Shevchenko V.,  2000, \mn@doi [\icarus] {10.1006/icar.2000.6410},
  147, 94

\bibitem[\protect\citeauthoryear{{Bowell}, {Hapke}, {Domingue}, {Lumme},
  {Peltoniemi}  \& {Harris}}{{Bowell} et~al.}{1989}]{bowellhg}
{Bowell} E.,  {Hapke} B.,  {Domingue} D.,  {Lumme} K.,  {Peltoniemi} J.,
  {Harris} A.~W.,  1989, in {Binzel} R.~P.,  {Gehrels} T.,   {Matthews} M.~S.,
  eds, Asteroids II. pp 524--556

\bibitem[\protect\citeauthoryear{Bradley et~al.,}{Bradley
  et~al.}{2020}]{photutils}
Bradley L.,  et~al., 2020, astropy/photutils: 1.0.0,
  \mn@doi{10.5281/zenodo.4044744}, \url
  {https://doi.org/10.5281/zenodo.4044744}

\bibitem[\protect\citeauthoryear{Brownlee et~al.,}{Brownlee
  et~al.}{2004}]{brownlee2004surface}
Brownlee D.~E.,  et~al., 2004, \mn@doi [Science] {10.1126/science.1097899},
  304, 1764

\bibitem[\protect\citeauthoryear{{Buratti}, {Hicks}, {Soderblom}, {Britt},
  {Oberst}  \& {Hillier}}{{Buratti} et~al.}{2004}]{buratti2004}
{Buratti} B.~J.,  {Hicks} M.~D.,  {Soderblom} L.~A.,  {Britt} D.,  {Oberst} J.,
    {Hillier} J.~K.,  2004, \mn@doi [\icarus] {10.1016/j.icarus.2003.05.002},
  \href {https://ui.adsabs.harvard.edu/abs/2004Icar..167...16B} {167, 16}

\bibitem[\protect\citeauthoryear{{Campins}, {A'Hearn}  \& {McFadden}}{{Campins}
  et~al.}{1987}]{campins1987}
{Campins} H.,  {A'Hearn} M.~F.,   {McFadden} L.-A.,  1987, \mn@doi [\apj]
  {10.1086/165249}, \href
  {https://ui.adsabs.harvard.edu/abs/1987ApJ...316..847C} {316, 847}

\bibitem[\protect\citeauthoryear{{Campins}, {Ziffer}, {Licandro},
  {Pinilla-Alonso}, {Fern{\'a}ndez}, {de Le{\'o}n}, {Moth{\'e}-Diniz}  \&
  {Binzel}}{{Campins} et~al.}{2006}]{campins2006nuclear}
{Campins} H.,  {Ziffer} J.,  {Licandro} J.,  {Pinilla-Alonso} N.,
  {Fern{\'a}ndez} Y.,  {de Le{\'o}n} J.,  {Moth{\'e}-Diniz} T.,   {Binzel}
  R.~P.,  2006, \mn@doi [\aj] {10.1086/506253}, \href
  {https://ui.adsabs.harvard.edu/abs/2006AJ....132.1346C} {132, 1346}

\bibitem[\protect\citeauthoryear{Campo~Bagatin, Alema{\~n}, Benavidez,
  P{\'e}rez-Molina  \& Richardson}{Campo~Bagatin
  et~al.}{2020}]{bagatin2020gravitational}
Campo~Bagatin A.,  Alema{\~n} R.~A.,  Benavidez P.~G.,  P{\'e}rez-Molina M.,
  Richardson D.~C.,  2020, \mn@doi [\icarus] {10.1016/j.icarus.2019.113603},
  339, 113603

\bibitem[\protect\citeauthoryear{{Chambers} et~al.,}{{Chambers}
  et~al.}{2016}]{chambersps1}
{Chambers} K.~C.,  et~al., 2016, arXiv e-prints, \href
  {https://ui.adsabs.harvard.edu/abs/2016arXiv161205560C} {p. arXiv:1612.05560}

\bibitem[\protect\citeauthoryear{{Davidsson} et~al.,}{{Davidsson}
  et~al.}{2016}]{davidsson2016}
{Davidsson} B.~J.~R.,  et~al., 2016, \mn@doi [\aap]
  {10.1051/0004-6361/201526968}, 592, A63

\bibitem[\protect\citeauthoryear{{Delahodde}, {Meech}, {Hainaut}  \&
  {Dotto}}{{Delahodde} et~al.}{2001}]{2001delahodde}
{Delahodde} C.~E.,  {Meech} K.~J.,  {Hainaut} O.~R.,   {Dotto} E.,  2001,
  \mn@doi [\aap] {10.1051/0004-6361:20011028}, \href
  {https://ui.adsabs.harvard.edu/abs/2001A&A...376..672D} {376, 672}

\bibitem[\protect\citeauthoryear{{Della Giustina} et~al.,}{{Della Giustina}
  et~al.}{2019}]{Dellagiustina2019}
{Della Giustina} D.~N.,  et~al., 2019, \mn@doi [Nature Astronomy]
  {10.1038/s41550-019-0731-1}, \href
  {https://ui.adsabs.harvard.edu/abs/2019NatAs...3..341D} {3, 341}

\bibitem[\protect\citeauthoryear{{Devog{\`e}le}, {Rivet}, {Tanga}, {Bendjoya},
  {Surdej}, {Bartczak}  \& {Hanus}}{{Devog{\`e}le} et~al.}{2015}]{devogele2015}
{Devog{\`e}le} M.,  {Rivet} J.~P.,  {Tanga} P.,  {Bendjoya} P.,  {Surdej} J.,
  {Bartczak} P.,   {Hanus} J.,  2015, \mn@doi [\mnras] {10.1093/mnras/stv1740},
  453, 2232

\bibitem[\protect\citeauthoryear{{Durech} et~al.,}{{Durech}
  et~al.}{2009}]{durech2009sparse}
{Durech} J.,  et~al., 2009, \mn@doi [\aap] {10.1051/0004-6361:200810393}, \href
  {https://ui.adsabs.harvard.edu/abs/2009A&A...493..291D} {493, 291}

\bibitem[\protect\citeauthoryear{{\v Durech}, {Sidorin}  \& {Kaasalainen}}{{\v
  Durech} et~al.}{2010}]{durechdamit}
{\v Durech} J.,  {Sidorin} V.,   {Kaasalainen} M.,  2010, \mn@doi [\aap]
  {10.1051/0004-6361/200912693}, 513, A46

\bibitem[\protect\citeauthoryear{{\v{D}}urech et~al.,}{{\v{D}}urech
  et~al.}{2012}]{durech2012analysis}
{\v{D}}urech J.,  et~al., 2012, \mn@doi [\aap] {10.1051/0004-6361/201219396},
  547, A10

\bibitem[\protect\citeauthoryear{{Duxbury}, {Newburn}  \& {Brownlee}}{{Duxbury}
  et~al.}{2004}]{2004_81p}
{Duxbury} T.~C.,  {Newburn} R.~L.,   {Brownlee} D.~E.,  2004, \mn@doi [Journal
  of Geophysical Research (Planets)] {10.1029/2004JE002316}, \href
  {https://ui.adsabs.harvard.edu/abs/2004JGRE..10912S02D} {109, E12S02}

\bibitem[\protect\citeauthoryear{{Fern{\'a}ndez} \& {Sosa}}{{Fern{\'a}ndez} \&
  {Sosa}}{2015}]{fernandez2015}
{Fern{\'a}ndez} J.~A.,  {Sosa} A.,  2015, \mn@doi [\planss]
  {10.1016/j.pss.2015.07.010}, \href
  {https://ui.adsabs.harvard.edu/abs/2015P&SS..118...14F} {118, 14}

\bibitem[\protect\citeauthoryear{Fern{\'a}ndez et~al.,}{Fern{\'a}ndez
  et~al.}{1999}]{fernandez1999inner}
Fern{\'a}ndez Y.~R.,  et~al., 1999, \mn@doi [\icarus] {10.1006/icar.1999.6127},
  140, 205

\bibitem[\protect\citeauthoryear{Fern{\'a}ndez et~al.,}{Fern{\'a}ndez
  et~al.}{2013}]{fernandez2013thermal}
Fern{\'a}ndez Y.,  et~al., 2013, \mn@doi [Icarus]
  {10.1016/j.icarus.2013.07.021}, 226, 1138

\bibitem[\protect\citeauthoryear{{Fornasier} et~al.,}{{Fornasier}
  et~al.}{2015}]{fornasier2015}
{Fornasier} S.,  et~al., 2015, \mn@doi [\aap] {10.1051/0004-6361/201525901},
  583, A30

\bibitem[\protect\citeauthoryear{{Giorgini} et~al.,}{{Giorgini}
  et~al.}{1996}]{horizons}
{Giorgini} J.~D.,  et~al., 1996, in AAS/Division for Planetary Sciences Meeting
  Abstracts \#28. p. 25.04

\bibitem[\protect\citeauthoryear{Gonz{\'a}lez-Solares
  et~al.,}{Gonz{\'a}lez-Solares et~al.}{2008}]{gonzalez2008initial}
Gonz{\'a}lez-Solares E.~A.,  et~al., 2008, \mn@doi [\mnras]
  {10.1111/j.1365-2966.2008.13399.x}, 388, 89

\bibitem[\protect\citeauthoryear{{Harmon}, {Nolan}, {Ostro}  \&
  {Campbell}}{{Harmon} et~al.}{2004}]{harmon2004comets}
{Harmon} J.~K.,  {Nolan} M.~C.,  {Ostro} S.~J.,   {Campbell} D.~B.,  2004, in
  {Festou} M.~C.,  {Keller} H.~U.,   {Weaver} H.~A.,  eds, , Comets II.
Univ. of Arizona, Tucson, p.~265

\bibitem[\protect\citeauthoryear{{Harmon}, {Nolan}, {Giorgini}  \&
  {Howell}}{{Harmon} et~al.}{2010}]{harmon2010radar}
{Harmon} J.~K.,  {Nolan} M.~C.,  {Giorgini} J.~D.,   {Howell} E.~S.,  2010,
  \mn@doi [\icarus] {10.1016/j.icarus.2009.12.026}, \href
  {https://ui.adsabs.harvard.edu/abs/2010Icar..207..499H} {207, 499}

\bibitem[\protect\citeauthoryear{Harris \& Warner}{Harris \&
  Warner}{2020}]{contactbrickary}
Harris A.,  Warner B.~D.,  2020, \mn@doi [\icarus]
  {10.1016/j.icarus.2019.113602}, 339, 113602

\bibitem[\protect\citeauthoryear{{Hartmann}, {Tholen}  \&
  {Cruikshank}}{{Hartmann} et~al.}{1987}]{hartmanndormant}
{Hartmann} W.~K.,  {Tholen} D.~J.,   {Cruikshank} D.~P.,  1987, \mn@doi
  [\icarus] {10.1016/0019-1035(87)90005-4}, \href
  {https://ui.adsabs.harvard.edu/abs/1987Icar...69...33H} {69, 33}

\bibitem[\protect\citeauthoryear{{Hergenrother} et~al.,}{{Hergenrother}
  et~al.}{2019}]{2019hergenrother}
{Hergenrother} C.~W.,  et~al., 2019, \mn@doi [Nature Communications]
  {10.1038/s41467-019-09213-x}, \href
  {https://ui.adsabs.harvard.edu/abs/2019NatCo..10.1291H} {10, 1291}

\bibitem[\protect\citeauthoryear{{Horner}, {M{\"u}ller}  \& {Lykawka}}{{Horner}
  et~al.}{2012}]{2012horner}
{Horner} J.,  {M{\"u}ller} T.~G.,   {Lykawka} P.~S.,  2012, \mn@doi [\mnras]
  {10.1111/j.1365-2966.2012.21067.x}, \href
  {https://ui.adsabs.harvard.edu/abs/2012MNRAS.423.2587H} {423, 2587}

\bibitem[\protect\citeauthoryear{Jeans}{Jeans}{1919}]{jeans1919problems}
Jeans J.,  1919, Problems of cosmogony and stellar dynamics.
University Press

\bibitem[\protect\citeauthoryear{{Jewitt} \& {Meech}}{{Jewitt} \&
  {Meech}}{1988}]{jewittmeech1988}
{Jewitt} D.~C.,  {Meech} K.~J.,  1988, \mn@doi [\apj] {10.1086/166351}, \href
  {https://ui.adsabs.harvard.edu/abs/1988ApJ...328..974J} {328, 974}

\bibitem[\protect\citeauthoryear{Jorda et~al.,}{Jorda
  et~al.}{2016}]{jorda2016global}
Jorda L.,  et~al., 2016, Icarus, 277, 257

\bibitem[\protect\citeauthoryear{Jutzi \& Asphaug}{Jutzi \&
  Asphaug}{2015}]{2015jutzi}
Jutzi M.,  Asphaug E.,  2015, \mn@doi [Science] {10.1126/science.aaa4747}, 348,
  1355

\bibitem[\protect\citeauthoryear{{Jutzi}, {Benz}, {Toliou}, {Morbidelli}  \&
  {Brasser}}{{Jutzi} et~al.}{2017}]{2017jutzi}
{Jutzi} M.,  {Benz} W.,  {Toliou} A.,  {Morbidelli} A.,   {Brasser} R.,  2017,
  \mn@doi [\aap] {10.1051/0004-6361/201628963}, 597, A61

\bibitem[\protect\citeauthoryear{{Kaasalainen} \& {Torppa}}{{Kaasalainen} \&
  {Torppa}}{2001}]{kaasa1}
{Kaasalainen} M.,  {Torppa} J.,  2001, \mn@doi [\icarus]
  {10.1006/icar.2001.6673}, \href
  {https://ui.adsabs.harvard.edu/abs/2001Icar..153...24K} {153, 24}

\bibitem[\protect\citeauthoryear{{Kaasalainen}, {Torppa}  \&
  {Muinonen}}{{Kaasalainen} et~al.}{2001}]{kaasa2}
{Kaasalainen} M.,  {Torppa} J.,   {Muinonen} K.,  2001, \mn@doi [\icarus]
  {10.1006/icar.2001.6674}, \href
  {https://ui.adsabs.harvard.edu/abs/2001Icar..153...37K} {153, 37}

\bibitem[\protect\citeauthoryear{Keller et~al.,}{Keller
  et~al.}{1986}]{keller1986first}
Keller H.~U.,  et~al., 1986, Nature, 321, 320

\bibitem[\protect\citeauthoryear{Kelley \& Lister}{Kelley \&
  Lister}{2019}]{calviacat}
Kelley M. S.~P.,  Lister T.,  2019, \mn@doi{10.5281/zenodo.2635840}, \url
  {https://github.com/mkelley/calviacat}

\bibitem[\protect\citeauthoryear{{Kokotanekova}}{{Kokotanekova}}{2018}]{2018rkphd}
{Kokotanekova} R.~D.,  2018, PhD thesis, Open University Milton Keynes, UK

\bibitem[\protect\citeauthoryear{{Kokotanekova} et~al.,}{{Kokotanekova}
  et~al.}{2017}]{2017kokotanekova}
{Kokotanekova} R.,  et~al., 2017, \mn@doi [\mnras] {10.1093/mnras/stx1716},
  \href {https://ui.adsabs.harvard.edu/abs/2017MNRAS.471.2974K} {471, 2974}

\bibitem[\protect\citeauthoryear{{Kokotanekova}, {Snodgrass}, {Lacerda},
  {Green}, {Nikolov}  \& {Bonev}}{{Kokotanekova}
  et~al.}{2018}]{2018kokotanekova}
{Kokotanekova} R.,  {Snodgrass} C.,  {Lacerda} P.,  {Green} S.~F.,  {Nikolov}
  P.,   {Bonev} T.,  2018, \mn@doi [\mnras] {10.1093/mnras/sty1529}, \href
  {https://ui.adsabs.harvard.edu/abs/2018MNRAS.479.4665K} {479, 4665}

\bibitem[\protect\citeauthoryear{{Kresak}}{{Kresak}}{1987}]{kresakdormant}
{Kresak} L.,  1987, \aap, \href
  {https://ui.adsabs.harvard.edu/abs/1987A&A...187..906K} {187, 906}

\bibitem[\protect\citeauthoryear{{Lamy} \& {Toth}}{{Lamy} \&
  {Toth}}{2009}]{lamy2009colors}
{Lamy} P.,  {Toth} I.,  2009, \mn@doi [\icarus] {10.1016/j.icarus.2009.01.030},
  \href {https://ui.adsabs.harvard.edu/abs/2009Icar..201..674L} {201, 674}

\bibitem[\protect\citeauthoryear{{Lamy}, {Toth}, {Fernandez}  \&
  {Weaver}}{{Lamy} et~al.}{2004}]{lamy2004sizes}
{Lamy} P.~L.,  {Toth} I.,  {Fernandez} Y.~R.,   {Weaver} H.~A.,  2004, in
  {Festou} M.~C.,  {Keller} H.~U.,   {Weaver} H.~A.,  eds, , Comets II.
Univ. of Arizona, Tucson, p.~223

\bibitem[\protect\citeauthoryear{Lowry, Duddy, Rozitis, Green, Fitzsimmons,
  Snodgrass, Hsieh  et~al.}{Lowry et~al.}{2012}]{lowry2012nucleus}
Lowry S.,  Duddy S.,  Rozitis B.,  Green S.~F.,  Fitzsimmons A.,  Snodgrass C.,
   Hsieh H.~H.,   et~al., 2012, \mn@doi [\aap] {10.1051/0004-6361/201220116},
  548, A12

\bibitem[\protect\citeauthoryear{{McNeill} et~al.,}{{McNeill}
  et~al.}{2018}]{mcneill2018}
{McNeill} A.,  et~al., 2018, \mn@doi [\aj] {10.3847/1538-3881/aaeb8c}, \href
  {https://ui.adsabs.harvard.edu/abs/2018AJ....156..282M} {156, 282}

\bibitem[\protect\citeauthoryear{Mommert et~al.,}{Mommert
  et~al.}{2020}]{Mommert_2020}
Mommert M.,  et~al., 2020, \mn@doi [The Planetary Science Journal]
  {10.3847/PSJ/ab8ae5}, 1, 12

\bibitem[\protect\citeauthoryear{Mottola et~al.,}{Mottola
  et~al.}{2014}]{mottola2014rotation}
Mottola S.,  et~al., 2014, \mn@doi [\aap] {10.1051/0004-6361/201424590}, 569,
  L2

\bibitem[\protect\citeauthoryear{Press, Teukolsky, Vetterling  \&
  Flannery}{Press et~al.}{1992}]{press1992numerical}
Press W.~H.,  Teukolsky S.~A.,  Vetterling W.~T.,   Flannery B.~P.,  1992, CUP,
  Cambridge

\bibitem[\protect\citeauthoryear{{Preusker} et~al.,}{{Preusker}
  et~al.}{2015}]{preusker2015shape}
{Preusker} F.,  et~al., 2015, \mn@doi [\aap] {10.1051/0004-6361/201526349},
  \href {https://ui.adsabs.harvard.edu/abs/2015A&A...583A..33P} {583, A33}

\bibitem[\protect\citeauthoryear{{Ro{\.z}ek} et~al.,}{{Ro{\.z}ek}
  et~al.}{2022}]{rozek2022}
{Ro{\.z}ek} A.,  et~al., 2022, \mn@doi [\mnras] {10.1093/mnras/stac1835}, \href
  {https://ui.adsabs.harvard.edu/abs/2022MNRAS.515.4551R} {515, 4551}

\bibitem[\protect\citeauthoryear{{Safrit}, {Steckloff}, {Bosh}, {Nesvorny},
  {Walsh}, {Brasser}  \& {Minton}}{{Safrit}
  et~al.}{2021}]{safrit2021subtorques}
{Safrit} T.~K.,  {Steckloff} J.~K.,  {Bosh} A.~S.,  {Nesvorny} D.,  {Walsh} K.,
   {Brasser} R.,   {Minton} D.~A.,  2021, \mn@doi [Planetary Science Journal]
  {10.3847/PSJ/abc9c8}, \href
  {https://ui.adsabs.harvard.edu/abs/2021PSJ.....2...14S} {2, 14}

\bibitem[\protect\citeauthoryear{{Samarasinha} \& {Mueller}}{{Samarasinha} \&
  {Mueller}}{2013}]{samarasinha2013}
{Samarasinha} N.~H.,  {Mueller} B. E.~A.,  2013, \mn@doi [\apjl]
  {10.1088/2041-8205/775/1/L10}, \href
  {https://ui.adsabs.harvard.edu/abs/2013ApJ...775L..10S} {775, L10}

\bibitem[\protect\citeauthoryear{{Samarasinha}, {Mueller}, {Belton}  \&
  {Jorda}}{{Samarasinha} et~al.}{2004}]{samarasinha2004rotation}
{Samarasinha} N.~H.,  {Mueller} B.~E.~A.,  {Belton} M.~J.~S.,   {Jorda} L.,
  2004, in {Festou} M.~C.,  {Keller} H.~U.,   {Weaver} H.~A.,  eds, , Comets
  II.
Univ. of Arizona, Tucson, p.~281

\bibitem[\protect\citeauthoryear{{Schleicher}, {Knight}, {Skiff}  \&
  {Bair}}{{Schleicher} et~al.}{2022}]{schleicherDPS}
{Schleicher} D.,  {Knight} M.,  {Skiff} B.,   {Bair} A.,  2022, in AAS/Division
  for Planetary Sciences Meeting Abstracts. p. 309.03

\bibitem[\protect\citeauthoryear{{Schwartz}, {Michel}, {Jutzi}, {Marchi},
  {Zhang}  \& {Richardson}}{{Schwartz} et~al.}{2018}]{2018schwartz}
{Schwartz} S.~R.,  {Michel} P.,  {Jutzi} M.,  {Marchi} S.,  {Zhang} Y.,
  {Richardson} D.~C.,  2018, \mn@doi [Nature Astronomy]
  {10.1038/s41550-018-0395-2}, \href
  {https://ui.adsabs.harvard.edu/abs/2018NatAs...2..379S} {2, 379}

\bibitem[\protect\citeauthoryear{{Shevchenko} et~al.,}{{Shevchenko}
  et~al.}{2012}]{shevchenko2012}
{Shevchenko} V.~G.,  et~al., 2012, \mn@doi [\icarus]
  {10.1016/j.icarus.2011.11.001}, \href
  {https://ui.adsabs.harvard.edu/abs/2012Icar..217..202S} {217, 202}

\bibitem[\protect\citeauthoryear{{Snodgrass}, {Fitzsimmons}, {Lowry}  \&
  {Weissman}}{{Snodgrass} et~al.}{2011}]{snodgrass2011size}
{Snodgrass} C.,  {Fitzsimmons} A.,  {Lowry} S.~C.,   {Weissman} P.,  2011,
  \mn@doi [\mnras] {10.1111/j.1365-2966.2011.18406.x}, \href
  {https://ui.adsabs.harvard.edu/abs/2011MNRAS.414..458S} {414, 458}

\bibitem[\protect\citeauthoryear{{Snodgrass}, {Feaga}, {Jones}, {Kueppers}  \&
  {Tubiana}}{{Snodgrass} et~al.}{2022}]{snodgrass22missions}
{Snodgrass} C.,  {Feaga} L.,  {Jones} G.~H.,  {Kueppers} M.,   {Tubiana} C.,
  2022, {Past and Future Comet Missions} (\mn@eprint {arXiv} {2208.08476})

\bibitem[\protect\citeauthoryear{{Stern} et~al.,}{{Stern}
  et~al.}{2019}]{stern2019arrokoth}
{Stern} S.~A.,  et~al., 2019, \mn@doi [Science] {10.1126/science.aaw9771},
  \href {https://ui.adsabs.harvard.edu/abs/2019Sci...364.9771S} {364, aaw9771}

\bibitem[\protect\citeauthoryear{{Tatsumi} et~al.,}{{Tatsumi}
  et~al.}{2020}]{tatsumi2020}
{Tatsumi} E.,  et~al., 2020, \mn@doi [\aap] {10.1051/0004-6361/201937096},
  \href {https://ui.adsabs.harvard.edu/abs/2020A&A...639A..83T} {639, A83}

\bibitem[\protect\citeauthoryear{{Thomas} et~al.,}{{Thomas}
  et~al.}{2013a}]{thomas20139p}
{Thomas} P.,  et~al., 2013a, \mn@doi [\icarus] {10.1016/j.icarus.2012.02.037},
  \href {https://ui.adsabs.harvard.edu/abs/2013Icar..222..453T} {222, 453}

\bibitem[\protect\citeauthoryear{{Thomas} et~al.,}{{Thomas}
  et~al.}{2013b}]{thomas2013shape}
{Thomas} P.~C.,  et~al., 2013b, \mn@doi [\icarus]
  {10.1016/j.icarus.2012.05.034}, \href
  {https://ui.adsabs.harvard.edu/abs/2013Icar..222..550T} {222, 550}

\bibitem[\protect\citeauthoryear{{Tonry} et~al.,}{{Tonry}
  et~al.}{2012}]{tonry2012pan}
{Tonry} J.~L.,  et~al., 2012, \mn@doi [\apj] {10.1088/0004-637X/750/2/99},
  \href {https://ui.adsabs.harvard.edu/abs/2012ApJ...750...99T} {750, 99}

\bibitem[\protect\citeauthoryear{{Virkki} et~al.,}{{Virkki}
  et~al.}{2022}]{virkkibinary}
{Virkki} A.~K.,  et~al., 2022, \mn@doi [The Planetary Science Journal]
  {10.3847/PSJ/ac8b72}, \href
  {https://ui.adsabs.harvard.edu/abs/2022PSJ.....3..222V} {3, 222}

\makeatother
\end{thebibliography}



\appendix


\bsp	
\label{lastpage}
\end{document}